\documentclass{aa}
\usepackage{graphicx}
\usepackage{txfonts}
\usepackage[normalem]{ulem}
\usepackage{color}
\usepackage{bm}
\usepackage{booktabs} 
\usepackage[]{xcolor}
\usepackage{comment}
\usepackage{amsmath}
\begin{document}

   \title{
    Fundamental stellar parameters of benchmark stars \\
   from CHARA interferometry}
   \subtitle{III. Giant and subgiant stars\thanks{Tables A.1.--A.7. are available at the CDS via anonymous ftp
to cdsarc.u-strasbg.fr (130.79.128.5)
or via http://cdsweb.u-strasbg.fr/cgi-bin/qcat?J/A+A/}}

 \author{I. Karovicova
          \inst{\ref{heidelberg}},
                    T. R. White\inst{\ref{sydney},\ref{aarhus}},
                    T. Nordlander\inst{\ref{rsaa},\ref{astro3d}},
                    L. Casagrande\inst{\ref{rsaa},\ref{astro3d}},
                    M. Ireland\inst{\ref{rsaa}},
                    D. Huber\inst{\ref{honolulu}}
          }
   \institute{Landessternwarte, University of Heidelberg
              K\"{o}nigstuhl 12, 69117, Heidelberg, Germany\\
              \email{karovicova@uni-heidelberg.de} \label{heidelberg}
                            \and 
        Sydney Institute for Astronomy (SIfA), School of Physics, University of Sydney, NSW 2006, Australia
         \label{sydney}
         \and
            Stellar Astrophysics Centre (SAC), Department of Physics and
            Astronomy, Aarhus University,
            Ny Munkegade 120, DK-8000 Aarhus~C, Denmark
        \label{aarhus}
        \and
                      Research School of Astronomy \& Astrophysics, Australian National University, Canberra, ACT 2611, Australia \label{rsaa}
                          \and
                      Center of Excellence for Astrophysics in Three Dimensions (ASTRO-3D), Australia \label{astro3d}
                                               \and  
          Institute for Astronomy, University of Hawai`i, 2680 Woodlawn Drive, Honolulu, HI 96822, USA \label{honolulu}
             }

   \date{Received August 23 2021 ; accepted October 12 2021}

  \abstract {Large spectroscopic surveys of the Milky Way must be calibrated against a sample of benchmark stars to ensure the reliable determination of atmospheric parameters.
  }
   {Here, we present new fundamental stellar parameters of seven giant and subgiant stars that will serve as benchmark stars for large surveys. The aim is to reach a precision of 1\% in the effective temperature. This precision is essential for accurate determinations of the full set of fundamental parameters and abundances for stars observed by the stellar surveys.}
     {We observed HD\,121370 ($\eta$\,Boo), HD\,161797 ($\mu$\,Her), HD\,175955, HD\,182736, HD\,185351, HD\,188512 ($\beta$\,Aql), and HD\,189349, using the high angular resolution optical interferometric instrument PAVO at the CHARA Array.
    The limb-darkening corrections were determined from 3D model atmospheres based on the STAGGER grid. The $T_\mathrm{eff}$ were determined directly from the Stefan-Boltzmann relation, with an iterative procedure to interpolate over tables of bolometric corrections.
   We estimated surface gravities from comparisons to Dartmouth stellar evolution model tracks. 
   The spectroscopic observations were collected from the ELODIE and FIES spectrographs. We estimated metallicities ($\mathrm{[Fe/H]}$) from a 1D non-local thermodynamic equilibrium (NLTE) abundance analysis of unblended lines of neutral and singly ionised iron.
     }
   {For six of the seven stars, we measured the value of $T_\mathrm{eff}$ to better than 1\% accuracy. For one star, HD\,189349, the uncertainty on $T_\mathrm{eff}$ is 2\%, due to an uncertain bolometric flux. We do not recommend this star as a benchmark until this measurement can be improved.
   Median uncertainties for all stars in $\log\,g$ and $\mathrm{[Fe/H]}$ are 0.034\,dex and 0.07\,dex, respectively.}
   {This study presents updated fundamental stellar parameters of seven giant and subgiant stars that can be used as a new set of benchmarks. All the fundamental stellar parameters were established on the basis of consistent combinations of interferometric observations, 3D limb-darkening
modelling, and spectroscopic analysis. This paper in this series follows our previous papers featuring dwarf stars
and stars in the metal-poor range.}

   
   \authorrunning{Karovicova et al.}
   \titlerunning{Fundamental stellar parameters of benchmark stars} 
   
   \maketitle
%

\section{Introduction}

Large stellar surveys are collecting a vast amount of data that allows us to investigate the properties of stars and planets and, thus, to explore our Galaxy in greater detail. These surveys include Gaia \citep{gaia16}, APOGEE \citep{alendeprieto08}, GALAH \citep{desilva15}, Gaia-ESO \citep{Gilmore12,Randich13}, and the future 4MOST survey \citep{dejong12}.
The quality of the information that can be extracted from the data observed by these surveys is strongly affected by the ability to properly test and correctly scale the stellar models applied to the surveys.
This objective is achieved with assistance of a sample of stars
with highly reliable fundamental stellar parameters, known as benchmarks stars.

It is crucial to determine the fundamental stellar parameters of benchmark stars with the highest possible precision. These parameters include the effective temperature ($T_\mathrm{eff}$), surface gravity 
($\log\,g$), metallicities ($\mathrm{[Fe/H]}$), and radius.
This is particularly important with regard to the effective temperature.
This parameter is traditionally determined by
using stellar spectroscopy. This technique unfortunately allows for only an indirect determination of 
$T_\mathrm{eff}$, leading to model dependencies that can negatively affect
the accuracy of delivered results. 
\citet{lebzelter12} points out the
difficulties, especially when  modelling giants, due to increased non-LTE effects and  to the differences between models.

For some stars, however, it is possible to directly determine $T_\mathrm{eff}$ with only minor model dependencies.
Optical interferometry offers this ideal approach by allowing for the precise measurement of the angular diameter, $\theta$ 
\citep[e.g.][]{Boyajian12a, Boyajian12b, Boyajian13, vonbraun14, Ligi16, baines18, rabus19, rains20}. From the angular diameter, in combination with the bolometric flux, $F_\mathrm{bol}$, the effective temperature, $T_\mathrm{eff}$, can be determined directly as described by the Stefan-Boltzmann relation. 
We note the effective temperatures determined in this matter are only weakly model-dependent via adopted bolometric and limb-darkening corrections. 
Stars with interferometrically derived $T_\mathrm{eff}$ are, therefore, ideal  benchmarks for calibrating spectroscopic surveys.
 
Currently, there is only a limited sample of interferometrically observed benchmark stars that has been defined. These are 34 stars selected for the Gaia-ESO survey \citep{Jofre14,heiter15}. 
However, even though the angular diameters of the Gaia-ESO benchmark stars were observed using optical interferometry and, therefore, the $T_\mathrm{eff}$ were directly determined, the measurements were selected from the literature, 
applying inconsistent limb-darkening corrections 
from various model atmosphere grids,
leading to strong inconsistencies within the sample. This set is rather limited
 coarsely samples stellar parameter space and, moreover, it is biased in terms of
the stellar age 
\citep{sahlholdt19}. 

We have established a program to expand the sample of benchmark stars. In general, we aim to both expand the sample and deliver highly reliable fundamental stellar parameters of stars covering a wide range of parameter space. 
Our sample includes stars that expand the parameter space of the current benchmark sample, as well as stars listed as Gaia-ESO benchmarks, for which we seek to confirm and revise their parameters, as well as stars that have been proposed as potential benchmarks by \citet{heiter15}.

As a first step within this programme, we observed three important Gaia-ESO metal-poor benchmark stars \citep{karovicova18}. We resolved previously reported differences between $T_\mathrm{eff}$ derived by spectroscopy, photometry, and interferometry pointed out by \citet{heiter15}. 
We further expanded the sample to include a set of ten metal-poor benchmark stars \citep{karovicova20}. Most recently, we determined the properties of nine dwarfs stars \citep{karovicova21}.
Overall, the published samples include updated stellar parameters for four Gaia-ESO benchmarks and for 15 entirely new benchmarks.
This comprises the subject of the next paper in this series, which is based on the investigation of seven giant and subgiant stars, including one of the Gaia-ESO benchmarks.
%
The recommended stars will serve as standards to validate current and future large stellar surveys as well as to help to calibrate the characterization of exoplanet host stars \citep{tayar20}.

\section{Observations}

\subsection{Science targets}

The third set of targets from our sample are seven 
giant and subgiant stars HD\,121370 ($\eta$\,Boo), HD\,161797 ($\mu\,$Her), HD\,175955, HD\,182736, HD\,185351, HD\,188512 ($\beta$\,Aql), HD\,189349. 
These stars were selected as possible candidates for benchmark stars that will be used for testing stellar models and validating large stellar surveys. 

\begin{figure}
    \includegraphics[width=.5\textwidth]{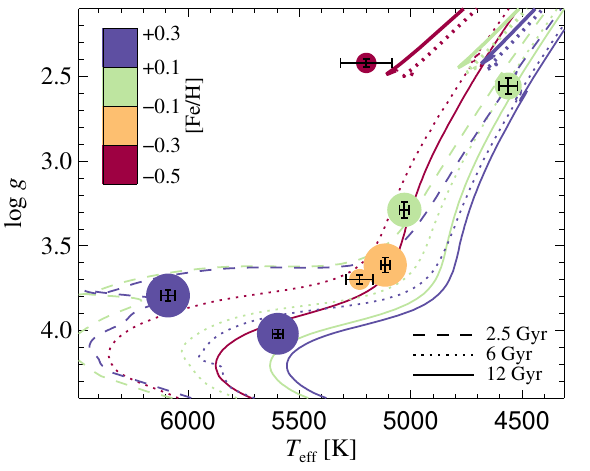}
    \caption{Stellar parameters of our program stars, colour-coded by metallicity, compared to theoretical Dartmouth isochrones of different ages (linestyles), and with metallicities of$\rm$ $\mathrm{[Fe/H]}$ = $+0.2$, $0.0$ and $-0.5$ (colours). Stellar evolution tracks for horizontal branch and AGB stars are overlaid based on the approximate RGB tip mass at each age and metallicity.
    Formal 1$\sigma$ uncertainties are shown by the error bars. The symbol size is logarithmically proportional to the angular diameter of each star.}
    \label{fig:HRdiag}
\end{figure}

All stars were selected with collaboration with spectroscopy teams of large stellar surveys. The stars have sizes and grades of brightness that allow us to reliably measure their angular diameters using the chosen interferometer and, thereby, to derive reliable values for $T_\mathrm{eff}$. 
These giant and subgiant stars will be added to the first set of metal-poor stars and second set of dwarf stars from this programme by \citet{karovicova20} and \citet{karovicova21}, respectively.
The metallicities of our stars range between $-0.6$ and $+0.3$\,dex.
The stellar parameters of the seven giant and subgiant stars are listed in Table~\ref{tab:parameters} and shown in Fig~\ref{fig:HRdiag}.

All the stars in our giant and subgiant sample have previous interferometric observations. 
We discuss these previous measurements and compare them with our new measurements in Sect.~\ref{comparison}.
One of the stars, namely, HD\,121370 ($\eta$\,Boo), is a current Gaia-ESO benchmark star. 
The star was interferometrically observed by \citet{nordgren01}, \citet{mozurkewich03}, \citet{thevenin05}, \citet{vanBelle07}, and \citet{baines14}. 
The Gaia-ESO benchmark sample adopted the angular diameter measured by \citet{vanBelle07}.
We are revising the measurement of this Gaia-ESO benchmark in this study.
We also present the measurements for five stars that have been proposed as future benchmarks in \citet{heiter15}.

Three targets are in multiple systems, that is:\ HD\,121370 ($\eta$\,Boo), HD\,161797 ($\mu$\,Her), and HD\,188512 ($\beta$\,Aql).
In particular, HD\,121370 ($\eta$\,Boo) is a long-period spectroscopic binary with a separation of more than 115\,arcsec  from a faint companion that does not contribute to the flux in PAVO \citep{mason01}.
Next, HD\,161797 ($\mu$\,Her) is a quadruplet system at a distance 8.3\,pc.
It consists of the G5IV primary and several M dwarfs \citep{roberts16}. The Ab component is separated by 1.8\,arcsec, and the BC pair by 35\,arcsec from the primary \citep{mason01}. All the components make a negligible contribution to the flux in PAVO.
Finally, HD\,188512 ($\beta$\,Aql) is a binary system. The companion,
$\beta$\,Aql\,B, is a V = 11.4\,mag M3 dwarf separated by approximately 13.4\,arcsec \citep{mason01} and it makes a negligible contribution to the flux.

\subsection{Interferometric observations and data reduction}\label{sec:observations}

We observed these stars using the Precision Astronomical Visible Observations (PAVO) interferometric instrument \citep{Ireland2008} at the CHARA (Center for High Angular Resolution Astronomy) Array at Mt. Wilson Observatory, California \citep{tenBrummelaar05}.
PAVO is operating as a pupil-plane beam combiner in optical wavelengths
between $\sim$\,630--880\,nm. 
The CHARA Array offers the longest available baselines in the optical wavelengths worldwide,
with baselines up to 330\,m.
The limiting magnitude of the observed targets is of
R\,$\sim$\,7.5. If the weather conditions allow,
the limiting magnitude can be slightly extended up to R\,$\sim$\,8.

The stars were observed using baselines between 65.91\,m and 278.5\,m. 
We collected the observations between 
2015 Apr 28
and 2018 Aug 7. Additionally, four of our targets have been previously observed with PAVO between 2009--2013.
These observations were reanalysed. This includes a new data reduction, so that all the presented data have been analysed consistently. This is mainly because in comparison to previous studies our treatment of limb-darkening differs.
The previously published data include the observations of HD\,175955, HD\,182736, HD\,189349 \citep{huber12}, and HD\,185351 \citep{johnson14}.
The dates 
of all observations, telescope configuration, and the baselines, $B$, are summarized in Table~\ref{table:2}.

For the data reduction, we used the PAVO reduction software, which has been broadly tested, especially for single baseline squared visibility ($V^2$) observations, and
 has been previously used in multiple studies \citep[e.g.][]{Bazot11,Derekas11,huber12,Maestro13}. 
Within the data reduction software, we applied several visibility corrections. 
This includes a coherence time ($t_0$) correction. This correction is based on a ratio of coherent and incoherent visibility estimators and may be used to correct visibility losses. However, since it can introduce biases, it was not used consistently throughout the whole study.
We note, however, that this correction may have been used in the previous studies of HD\,175955, HD\,182736, HD\,189349 \citep{huber12}, and HD\,185351 \citep{johnson14}.

Additionally, we also changed the used wavelength range. In our study, we include all 38 channels, which range between 630 and 880\,nm. In the previous studies, only the central 23 channels were used, ranging between 650 and 800\,nm. The shorter range was used to avoid the possibly unreliable data towards the end of the range.
We found our data was internally consistent within uncertainties, even though the edges of the bandpass typically had higher uncertainties, and, therefore, we included the full range. The absolute uncertainty on the wavelength scale was set to 5\,nm.

In order to monitor the interferometric transfer function,
a set of calibrating stars was observed
immediately before and after the science targets.
We selected the calibrating stars from the CHARA catalog of calibrators 
and from the Hipparcos catalogue \citep{hipparcos}. 
The calibrating stars were selected to be either unresolved or nearly unresolved sources
located close on sky to our science targets. 
We determined the angular diameters of the calibrators using the $V-K$ relation of \citet{boyajian14}. We corrected for limb-darkening to determine the uniform disc diameter in the $R$ band.

We used $V$-band magnitudes from the Tycho-2 catalogue \citep{Hoeg2000}. We converted them into the Johnson system using the calibration by \citet{bessell00}.
We selected the $K$-band magnitudes from the Two Micron All Sky Survey \citep[2MASS;][]{Skrutskie2006}.
We estimated the reddening from the dust map of \citet{green15} and we applied the reddening law of \citet{odonnell94}.
The relative uncertainty on calibrator diameters was set to 5\% \citep{boyajian14}, covering both the uncertainty on the calibrator diameters as well as the reddening.

All the calibrating stars were checked for binarity
in Gaia\,DR2, the proper motion anomaly \citep{kervella19}, the $\rm{\tt phot\_bp\_rp\_excess\_factor}$ \citep{evans18}, and the renormalised unit weight error \citep[RUWE;][]{belokurov20}. All these sources suggest that none of our calibrators have a companion that
is large enough to affect our interferometric measurements or estimated calibrator sizes.
 
Since we noticed that using 
high-order limb-darkening coefficients to estimate of the calibrator sizes 
has a negligible impact ($\sim$0.3\%) that is significantly smaller than our 5\% 
uncertainty in the calibrator diameters, the angular sizes of the calibrators were estimated as a uniform disc in the $R$-band.
The calibrating stars and
their spectral type, magnitude in the $V$ and $R$ band, their expected angular diameter,
and the corresponding science targets are summarized in Table~\ref{table:3}.

        \begin{table*}
  \caption{Stellar parameters}    
  \centering   
  \begin{tabular}{l l l l l l l r r }      
\hline\hline   
   Star      && Sp.type  & Right ascension &    Declination &    m$_V$ &    m$_R$ & E(B-V)&   $\pi$      \\ 
   &&& & & (mag) & (mag) & (mag)&   \\
     \hline    
HD\,121370   &$\eta$ Boo& G0IV& 13 54 41.0789 &  $+$18 23 51.7946 & 2.68 & 2.24 &   0     $\pm$ 0    & 87.750 $\pm$ 1.240 \\
HD\,161797   &$\mu$ Her&G5IV & 17 46 27.5267 &  $+$27 43 14.4380 & 3.42 & 2.89 &   0     $\pm$ 0    & 119.110 $\pm$ 0.480 \\
HD\,175955   && K0III & 18 55 33.3265 &  $+$47 26 26.7847 & 7.02 & 7.02 &   0.01 $\pm$ 0.02 & 7.366 $\pm$ 0.030  \\
HD\,182736  && G8IV & 19 24 03.3794 &  $+$44 56 00.7352 & 7.01 & 7.02 &   0     $\pm$ 0    & 17.176 $\pm$ 0.025 \\
HD\,185351  && G8.5III & 19 36 37.9763 &  $+$44 41 41.7600 & 5.17 & 5.18 &   0     $\pm$ 0    & 24.225 $\pm$ 0.082 \\
HD\,188512  &$\beta$ Aql& G8IV & 19 55 18.7926 &  $+$06 24 24.3425 & 3.71 & 3.05 &   0     $\pm$ 0    & 74.760 $\pm$ 0.360 \\
HD\,189349  && G4III-IV & 19 58 02.3822 &  $+$40 55 36.6322 & 7.31 & 7.32 &   0.02 $\pm$ 0.03 & 4.964 $\pm$ 0.340  \\
  \hline                                 
\end{tabular}
\tablefoot{Parallaxes are from Gaia DR2, without any zero point correction. }
\label{tab:parameters}
  \end{table*}

\begin{table*}[t!]
\caption{Interferometric observations: Giants and subgiants}            
\label{table:2}      
\centering                          
\begin{tabular}{l l l c c l }      
\hline\hline            
Science target & UT date & Telescope & $B$ (m) ) & \# of obs. & Calibrator stars\\  
\hline 
   HD 121370  & 2015 Apr 28  & E1E2  &  65.91   & 4 & HD 116706, HD 121996 \\
              & 2015 Apr 29  & E1E2  &  65.91   & 4 & HD 116706, HD 121996 \\    
              & 2015 Apr 30  & E1E2  &  65.91   & 3 & HD 116706, HD 121996 \\       
              & 2018 Aug 5   & E1E2  &  65.91   & 2 & HD 116706, HD 121996 \\   
              & 2018 Aug 7   & E1E2  &  65.91   & 2 & HD 121996 \\                 
   HD 161797  & 2015 Apr 28  & E1E2  &  65.91   & 2 & HD 157087, HD 166230 \\ 
              & 2015 Apr 29  & E1E2  &  65.91   & 2 & HD 157087, HD 166230 \\         
              & 2015 Jun 18  & E1E2  &  65.91   & 2 & HD 166230 \\             
   HD 175955  & 2009 Jul 15  & S2W2  &  177.45  & 2 & HD 179095, HD 180501, HD 184147 \\
              & 2011 Jul 3   & W1W2  &  107.93  & 2 & HD 174177 \\               
   HD 182736  & 2009 Jul 15  & S2W2  &  177.45  & 1 & HD 183142, HD 188461 \\  
              & 2009 Jul 19  & S2E2  &  248.13  & 1 & HD 184147, HD 188461 \\     
              & 2010 Jul 20  & S2E2  &  248.13  & 1 & HD 179483 \\   
              & 2011 Jul 2   & S1W1  &  278.50  & 2 & HD 180681, HD 183142 \\                
   HD 185351  & 2013 Jul 7   & W1W2  &  107.93  & 5 & HD 177003, HD 185872, HD 188252 \\  
              & 2014 Apr 6   & W1W2  &  107.93  & 1 & HD 185872 \\     
              & 2014 Apr 7   & W1W2  &  107.93  & 2 & HD 177003, HD 185872 \\  
              & 2014 Apr 10  & E2W2  &  156.26  & 1 & HD 184787, HD 188252 \\           
   HD 188512  & 2015 Jul 27  & E1E2  &   65.91  & 3 & HD 179761, HD 188385 \\           
              & 2016 Aug 15  & E2W2  &  156.26  & 1 & HD 186689 \\           
              & 2016 Aug 17  & W1W2  &  107.93  & 4 & HD 186689, HD 191263 \\              
   HD 189349  & 2009 Jul 14  & S2W2  &  177.45  & 2 & HD 188461, HD 190025 \\ 
              & 2009 Jul 15  & S2W2  &  177.45  & 1 & HD 188461 \\      
              & 2011 Jul 5   & S1W2  &  210.98  & 1 & HD 189845 \\ 
\hline                                 
\end{tabular}
\end{table*}

   \begin{table}[t!]
\caption{Calibrator stars used for interferometric observations:\ Giants and subgiants}            
\label{table:3}      
\centering                          
\begin{tabular}{l l c c c l}      
\hline\hline            
Calibrator & Spectral & m$_V$ & m$_K$ & UD  \\ 
&type&&&(mas)\\
\hline 
 HD 116706  & A3IV  & 5.75 & 5.50 & 0.271 \\ 
 HD 121996  & A0V   & 5.75 & 5.70 & 0.239 \\ 
 HD 157087  & A3III & 5.36 & 5.14 & 0.320 \\ 
 HD 166230  & A8III & 5.10 & 4.61 & 0.419 \\ 
 HD 174177  & A2IV  & 6.51 & 6.26 & 0.191 \\ 
 HD 177003  & B2.5IV& 5.38 & 5.90 & 0.204 \\ 
 HD 179095  & A0    & 6.92 & 6.99 & 0.131 \\ 
 HD 179483  & A2V   & 7.20 & 6.90 & 0.143 \\ 
 HD 179761  & B7III & 5.15 & 5.28 & 0.271 \\ 
 HD 180501  & A0V   & 7.42 & 7.27 & 0.118 \\ 
 HD 180681  & A0V   & 7.48 & 7.39 & 0.111 \\ 
 HD 183142  & B8V   & 7.07 & 7.53 & 0.096 \\ 
 HD 184147  & B9IV  & 7.17 & 7.16 & 0.122 \\ 
 HD 185872  & B9III & 5.40 & 5.48 & 0.268 \\ 
 HD 186689  & A3IV  & 5.91 & 5.46 & 0.284 \\ 
 HD 188252  & B2III & 5.90 & 6.36 & 0.165 \\ 
 HD 188385  & A2V   & 6.13 & 6.00 & 0.213 \\ 
 HD 188461  & B2IV  & 6.99 & 7.45 & 0.098 \\ 
 HD 189845  & A0    & 7.22 & 7.08 & 0.128 \\ 
 HD 190025  & B5V   & 7.53 & 7.78 & 0.087 \\ 
 HD 191263  & B3V   & 6.34 & 6.72 & 0.142 \\ 
\hline                                 
\end{tabular}
\end{table}

\begin{table}\small
\caption{Angular diameters and linear limb-darkening coefficients.}   \label{table:linear}      
\centering                          
\begin{tabular}{llll}      
\hline\hline            
Star  & $\theta_\mathrm{UD}$ (mas) & \multicolumn{2}{l}{Linear limb-darkening$^a$}   \\   
     &                              &    $u$    &    $\theta_\mathrm{LD}$ (mas)   \\
\hline    

HD\,121370    & 2.088 $\pm$ 0.015 & 0.551 $\pm$ 0.012 & 2.206 $\pm$ 0.018 \\
HD\,161797    &  1.797 $\pm$ 0.015 & 0.605 $\pm$ 0.013 & 1.913 $\pm$ 0.016 \\
HD\,175955    &  0.632 $\pm$ 0.009 & 0.706 $\pm$ 0.010 & 0.678 $\pm$ 0.010 \\ 
HD\,182736    &  0.412 $\pm$ 0.004 & 0.621 $\pm$ 0.012 & 0.438 $\pm$ 0.004 \\
HD\,185351    &  1.057 $\pm$ 0.009 & 0.652 $\pm$ 0.011 & 1.129 $\pm$ 0.009 \\
HD\,188512    &  1.979 $\pm$ 0.018 & 0.633 $\pm$ 0.012 & 2.125 $\pm$ 0.016 \\
HD\,189349    &  0.401 $\pm$ 0.005 & 0.596 $\pm$ 0.013 & 0.423 $\pm$ 0.005 \\

\hline                                 
\end{tabular}
\flushleft $^{a}$ Limb-darkening coefficients derived from the grid of \citet{claret11}; see text for details. The final limb-darkened diameters using higher order limb-darkening model are listed in Table 5.
\end{table}

     \begin{table*}
   \caption{Observed ($\varTheta_{LD}$) and derived ($F_\mathrm{bol}$ , $M$, $L$, $R$) stellar parameters}    
   \centering   
  \begin{tabular}{l r r r r r}      
\hline\hline   
 Star  &     $\varTheta_{LD}$  &  $F_\mathrm{bol}$ &      $M$(M$_\odot$)    &   $L$ (L$_\odot$) &    $R$ (R$_\odot$)\\
   &  (mas)  & (erg s$^{-1}$cm$^{-2}$10$^{-8}$) & &  \\
    \hline     
HD\,121370  & 2.173 $\pm$ 0.018  &   216.552 $\pm$ 2.065 & 1.61 $\pm$ 0.11 & 8.758 $\pm$ 0.261  & 2.659 $\pm$ 0.044   \\
HD\,161797  & 1.888 $\pm$ 0.014  &   116.485 $\pm$ 0.721 & 1.11 $\pm$ 0.06 & 2.557 $\pm$ 0.026  & 1.704 $\pm$ 0.016   \\
HD\,175955  & 0.661 $\pm$ 0.009  &   6.345 $\pm$ 0.156   & 1.23 $\pm$ 0.12 & 36.417 $\pm$ 0.941 & 9.796 $\pm$ 0.137   \\
HD\,182736  & 0.433 $\pm$ 0.009  &   4.672 $\pm$ 0.102   & 1.34 $\pm$ 0.05 & 4.932 $\pm$ 0.109  & 2.711 $\pm$ 0.019   \\
HD\,185351  & 1.113 $\pm$ 0.009  &   26.396 $\pm$ 0.175  & 1.73 $\pm$ 0.17 & 14.008 $\pm$ 0.133 & 4.946 $\pm$ 0.043   \\
HD\,188512  & 2.096 $\pm$ 0.014  &   100.053 $\pm$ 0.738 & 1.36 $\pm$ 0.13 & 5.575 $\pm$ 0.068  & 3.012 $\pm$  0.025  \\
HD\,189349  & 0.417 $\pm$ 0.005  &   4.242 $\pm$ 0.366   & 0.73 $\pm$ 0.03 & 53.608 $\pm$ 4.686 & 9.084 $\pm$ 0.125   \\

  \hline                                 
\end{tabular}
\tablefoot{ $F_\mathrm{bol}$ and $L$ are obtained adopting L$_\odot=3.842 \times 10^{33}$ erg s$^{-1}$.}
\label{der_parameters}
  \end{table*}


{
\begin{sidewaystable*}
   \caption{Bolometric corrections}    
  \centering   
  \begin{tabular}{l l l l l l l l l l l l l l l l l l l}      
\hline\hline   
 & & & & & & & & & & & & & & & & & & \\
 Star        &     BC$_{H_p}$ &  BC$_{B_T}$ &  BC$_{V_T}$ &  BC$_J$ &  BC$_H$
      &  BC$_K$ &   $B_T$  &    e$B_T$    &  $V_T$   &  e$V_T$   &          $H_p$     & e$H_p$   &     $J$  &    e$J$  &    $H$  &    e$H$  &    $K$  &   e$K$ \\

 & & & & & & & & & & & & & & & & & & \\

 \hline

 & & & & & & & & & & & & & & & & & & \\
HD121370   &     -0.115 & -0.697 & -0.055 & 1.016 &  1.234 & 1.318 &  3.380 &   0.014 &  2.719 &  0.009     &  2.7957 & 0.0004 &    -   &   -    &   -   &   -   &    -   &    -    \\ 
 & & & & & & & & & & & & & & & & & & \\
HD161797   &     -0.207 & -0.984 & -0.146 & 1.180 &  1.481 & 1.580 &  4.336 &   0.014 &  3.490 &  0.009     &  3.5596 & 0.0005 &    -   &   -    &   -   &   -   &    -   &    -    \\
 & & & & & & & & & & & & & & & & & & \\
HD175955   &     -0.648 & -1.982 & -0.618 & 1.482 &  2.056 & 2.209 &  8.513 &   0.016 &  7.146 &  0.010     &  7.1808 & 0.0010 &  4.999 & 0.024  &   -   &   -   &  4.318 &  0.017  \\
 & & & & & & & & & & & & & & & & & & \\
HD182736   &     -0.313 & -1.169 & -0.259 & 1.276 &  1.673 & 1.775 &  8.021 &   0.016 &  7.103 &  0.010     &  7.1673 & 0.0009 &  5.515 & 0.024  &   -   &   -   &  5.028 &  0.016  \\
 & & & & & & & & & & & & & & & & & & \\
HD185351   &     -0.372 & -1.369 & -0.316 & 1.345 &  1.775 & 1.893 &  6.350 &   0.014 &  5.273 &  0.009     &  5.3355 & 0.0005 &    -   &   -    &   -   &   -   &    -   &    -    \\
 & & & & & & & & & & & & & & & & & & \\
HD188512   &     -0.350 & -1.257 & -0.298 & 1.312 &  1.736 & 1.844 &  4.789 &   0.014 &  3.807 &  0.009     &  3.8668 & 0.0004 &    -   &   -    &   -   &   -   &    -   &    -    \\
 & & & & & & & & & & & & & & & & & & \\
HD189349   &     -0.407 & -1.229 & -0.358 & 1.251 &  1.675 & 1.778 &  8.411 &   0.016 &  7.411 &  0.010     &  7.4738 & 0.0011 &  5.638 & 0.024  & 5.181 & 0.021 &  5.124 &  0.029  \\

 & & & & & & & & & & & & & & & & & & \\

   \hline  
\end{tabular}
\tablefoot{Adopted bolometric corrections (BC). Hipparcos $H_p$ and Tycho2
       $B_TV_T$ magnitudes for all stars. 2MASS $JHK_S$ only if the quality is
       flag `A' The zero-point of these bolometric corrections is set by $M_{bol,\odot}=4.75$. $F_\mathrm{bol}$ for each star is obtained from Eq (3) of \cite{Casagrande_VandenBerg18a}, adopting L$_{\odot}=3.842 \times 10^{33}$ erg s$^{-1}$.}
\label{tab:bolcor}
  \end{sidewaystable*}
}


     \begin{table}
   \caption{Derived stellar parameters ($T_\mathrm{eff}$, $\log\,g$, $[\mathrm{Fe/H}]$)}    
  \begin{tabular}{l l l  r }      
\hline\hline   
 Star &   $T_\mathrm{eff}$  &    $\log\,g$ &     $[\mathrm{Fe/H}]$  \\
    &      (K) & (dex)& (dex) \\ 
  \hline 
HD\,121370 & 6090 $\pm$ 29  & 3.794 $\pm$ 0.034 & 0.29 $\pm$ 0.07     \\
HD\,161797 & 5596 $\pm$ 22  & 4.020 $\pm$ 0.025 & 0.26 $\pm$ 0.04     \\
HD\,175955 & 4568 $\pm$ 42  & 2.556 $\pm$ 0.048 & 0.00 $\pm$ 0.13   \\
HD\,182736 & 5229 $\pm$ 37  & 3.699 $\pm$ 0.025 & $-$0.18 $\pm$ 0.04    \\
HD\,185351 & 5025 $\pm$ 22  & 3.288 $\pm$ 0.046 & $-$0.02 $\pm$ 0.07    \\ 
HD\,188512 & 5113 $\pm$ 20  & 3.614 $\pm$ 0.044 & $-$0.20 $\pm$ 0.04   \\ 
HD\,189349 & 5199 $\pm$ 116 & 2.388 $\pm$ 0.009 & $-$0.57 $\pm$ 0.09    \\
   \hline  
\end{tabular}
\label{der_parameters2}
  \end{table}

      \begin{figure}
   \centering
   \includegraphics[width=\hsize]{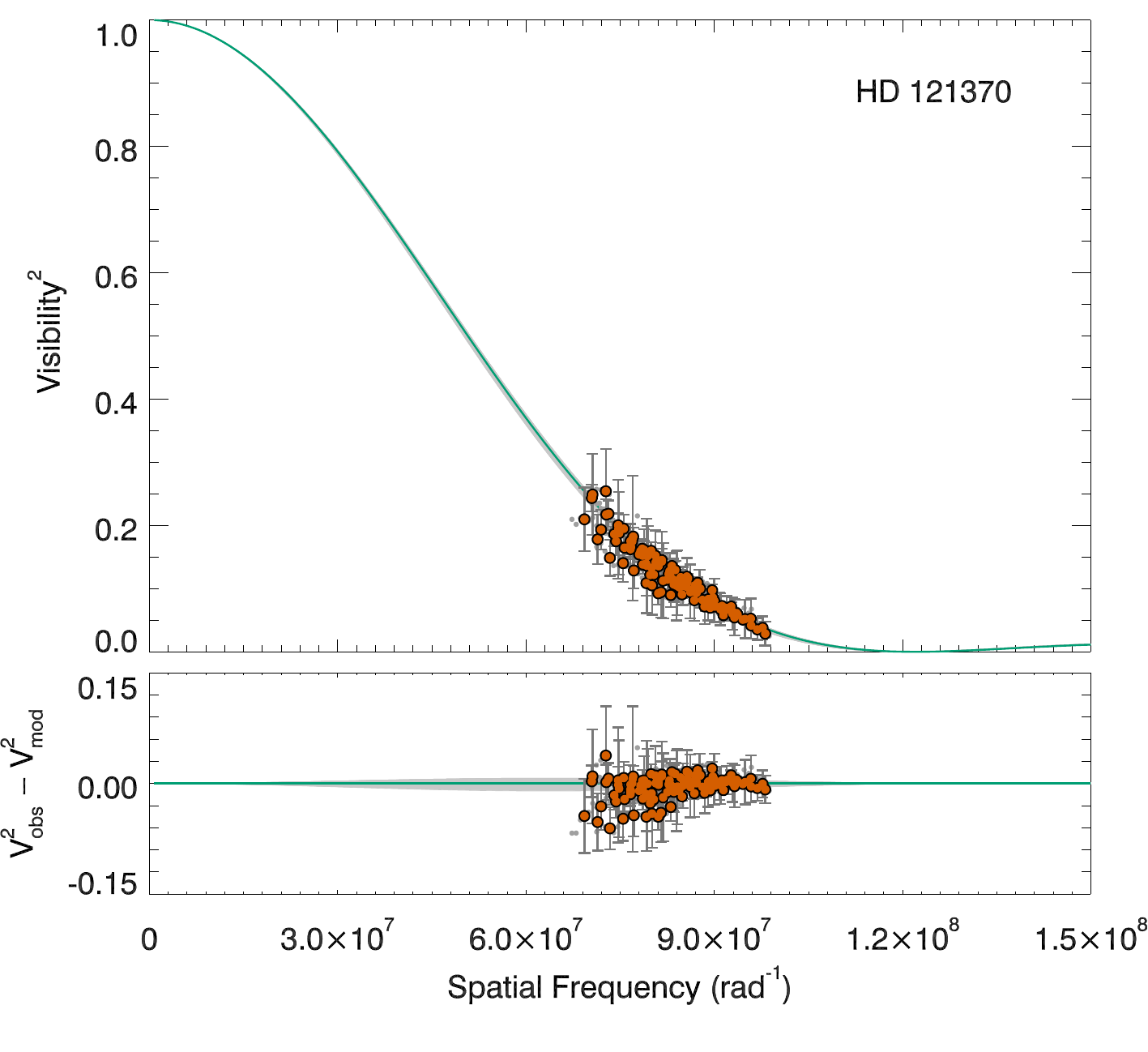} 
   \includegraphics[width=\hsize]{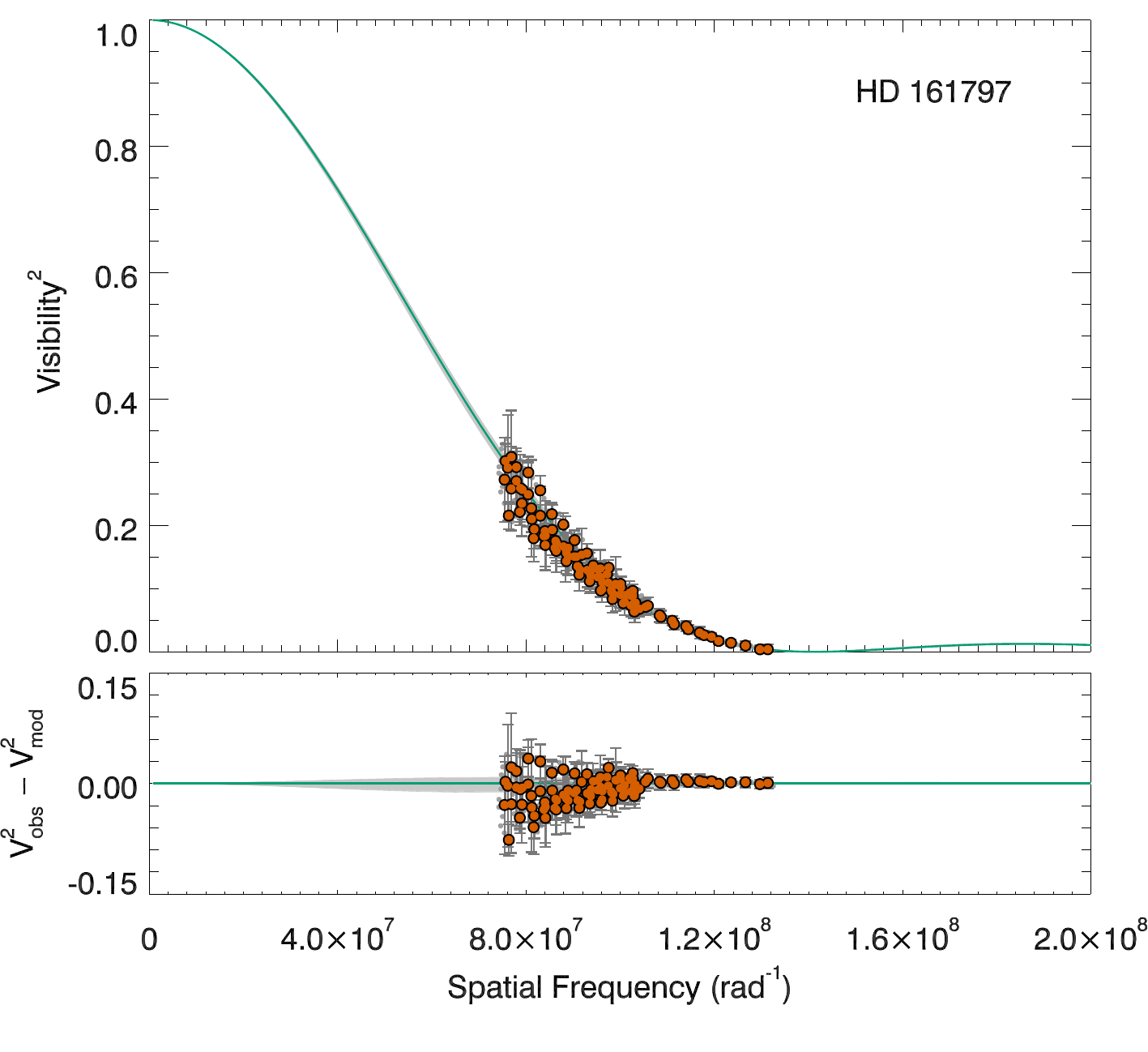} 

      \caption{Squared visibility versus spatial frequency for HD\,121370 and HD\,161797. The HD\, number is noted in the right upper corner in the each plot.
      The raw error bars have been scaled to the reduced $\chi^{2}$ before the final fitting. For HD\,121370, the reduced $\chi^{2}$\,=\,13.1 
      and for HD\,161797, 
      $\chi^{2}$\,=\,5.9.
      The grey dots are the individual PAVO measurements in each wavelength channel. 
      For clarity, we show weighted averages of the PAVO measurements 
      as red circles. The green line shows the fitted limb-darkened model to the PAVO data, with the light grey-shaded region indicating the 1-$\sigma$ uncertainties. The lower
      panel shows the residuals from the fit.}
      \label{figures1}
   \end{figure}
   
  
      \begin{figure}
   \centering

   \includegraphics[width=\hsize]{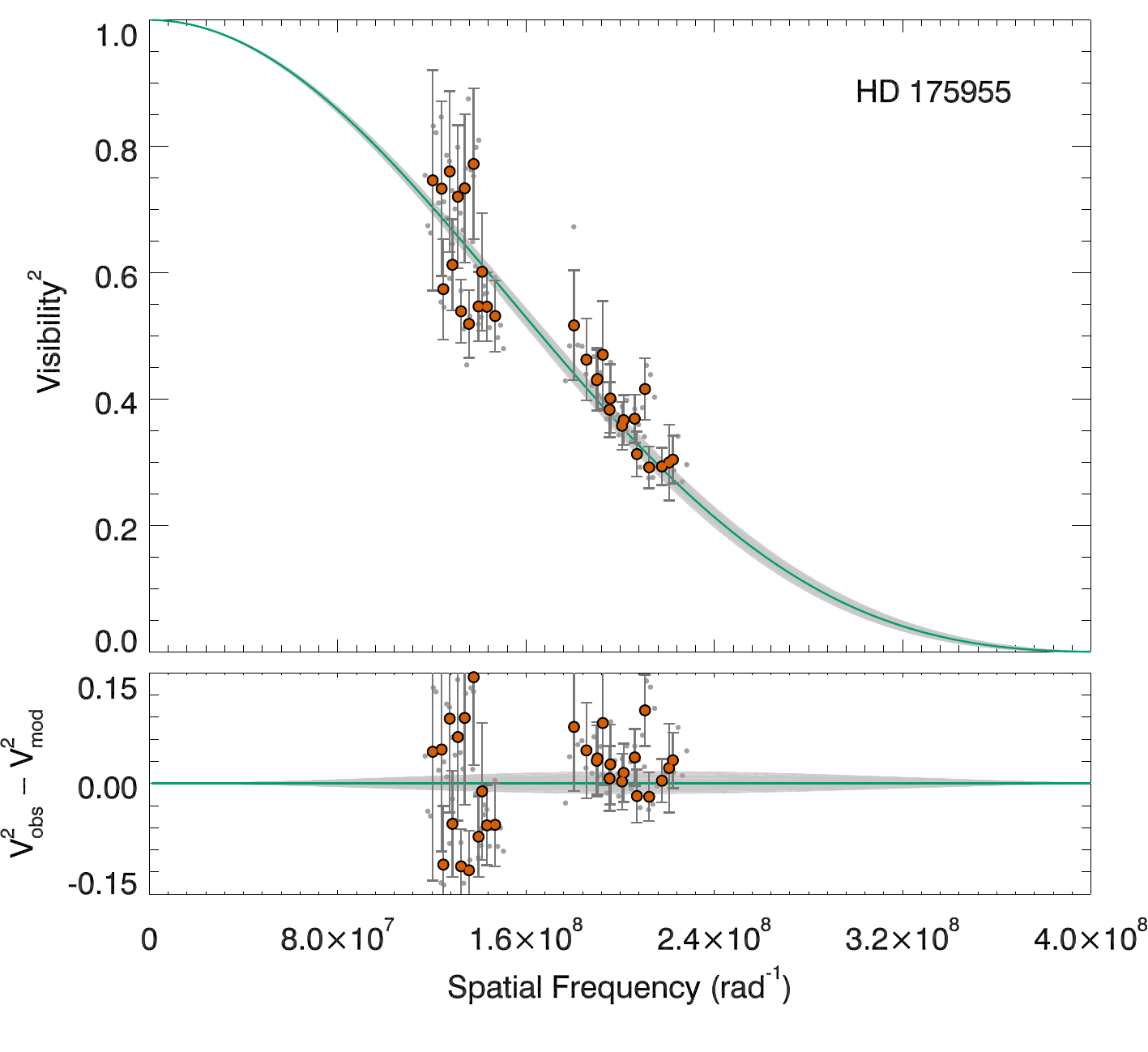}
   \includegraphics[width=\hsize]{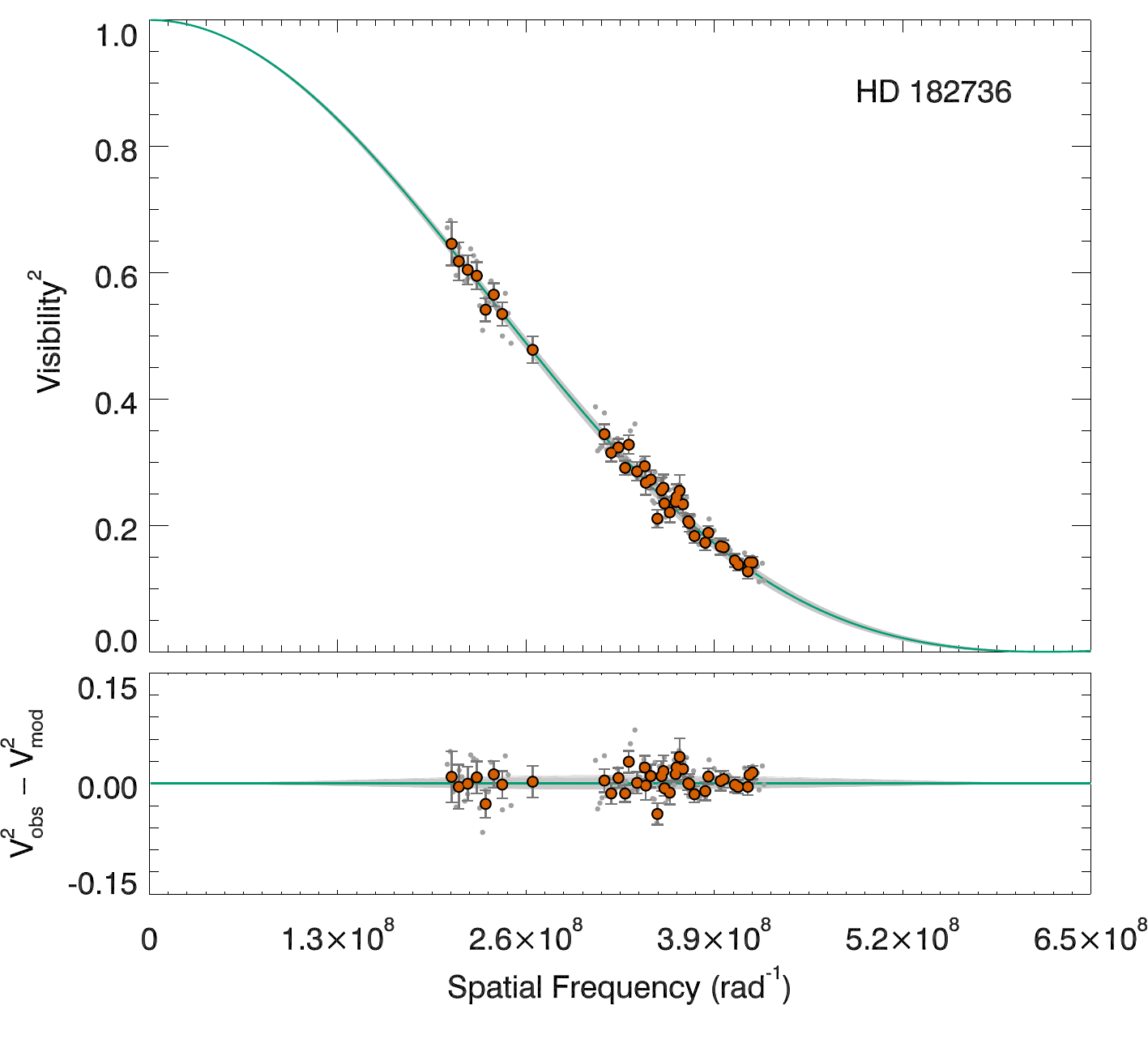}    
      \caption{Squared visibility vs. spatial frequency for HD\,175955 and
HD\,182736. The lower panel shows the residuals from the fit.
The error bars have been scaled to the reduced $\chi^{2}$. For HD\,175955
the reduced $\chi^{2}$\,=\,2.3 
and for HD\,182736
$\chi^{2}$\,=\,1.1. 
All lines and
symbols are the same as for Fig.\ref{figures1}. }
         \label{figures2}
   \end{figure}

      \begin{figure}
   \centering
   \includegraphics[width=\hsize]{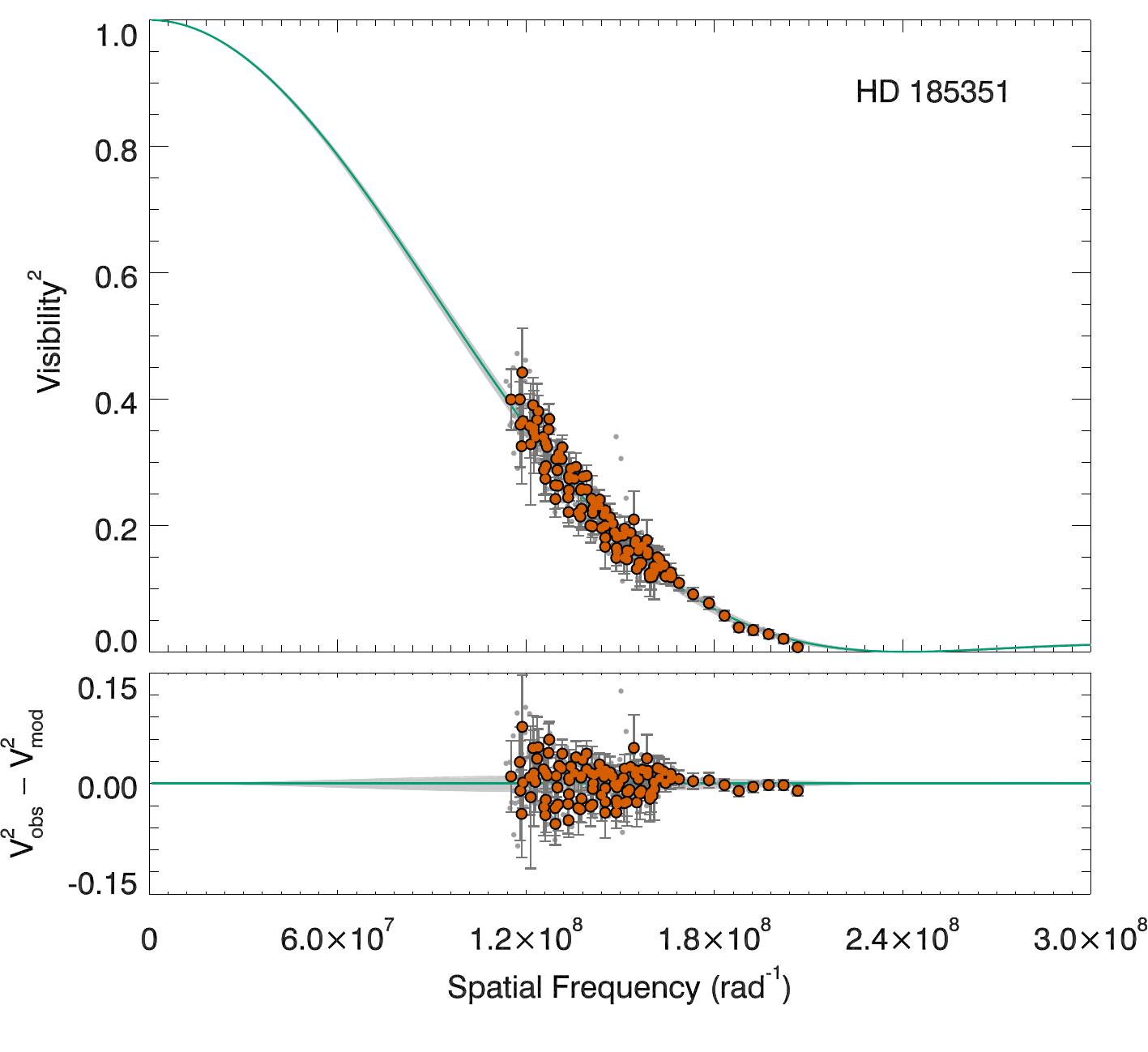}   
      \caption{Squared visibility vs. spatial frequency for HD\,185351. The lower panel shows the residuals from the fit.
The error bars have been scaled to the reduced $\chi^{2}$. The reduced $\chi^{2}$ for HD\,185351 is $\chi^{2}$\,=\,4.8. 
All lines and
symbols are the same as for Fig. \ref{figures1}. }
         \label{figures3}
   \end{figure}

      \begin{figure}
   \centering
    \includegraphics[width=\hsize]{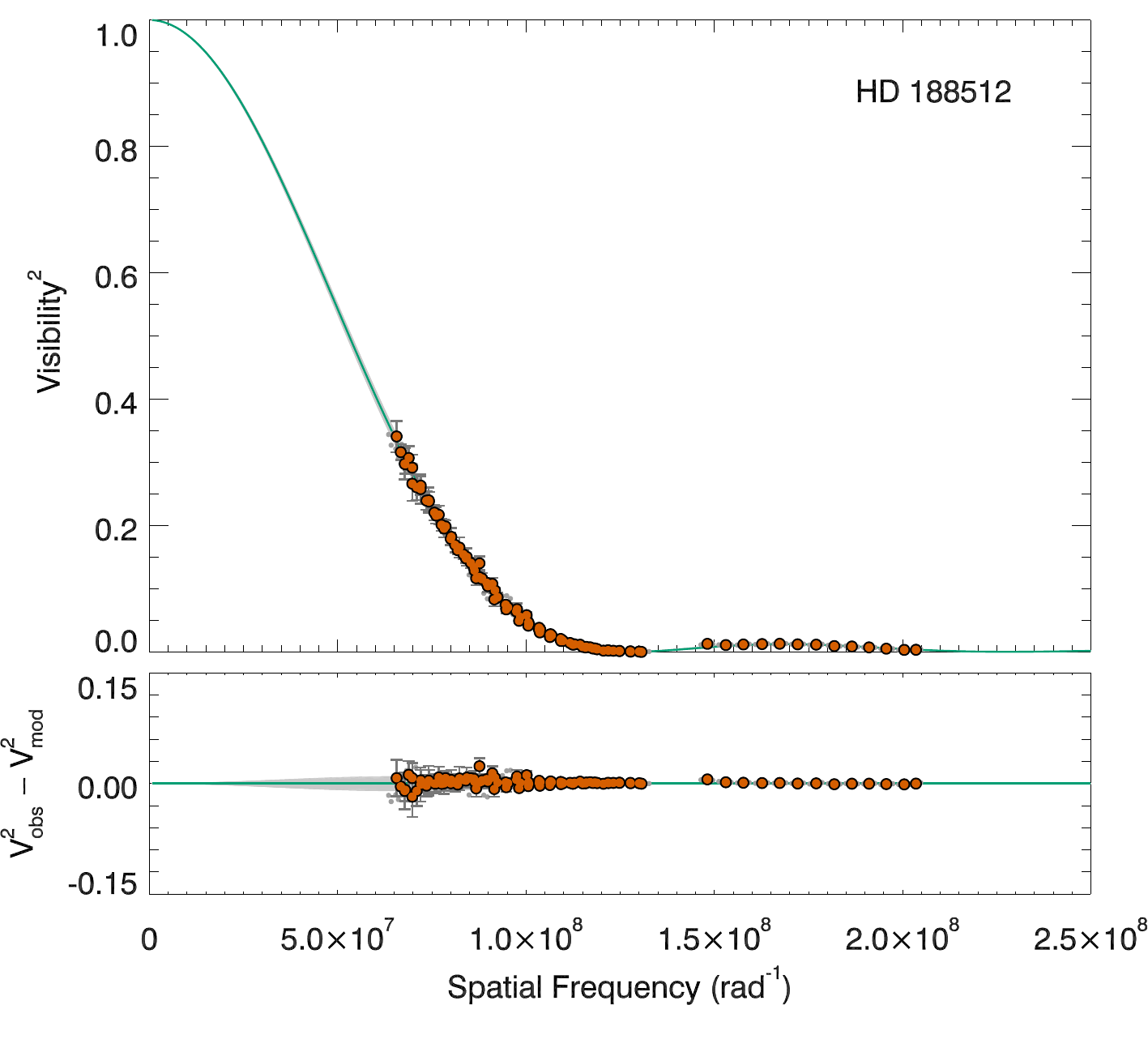}    
    \includegraphics[width=\hsize]{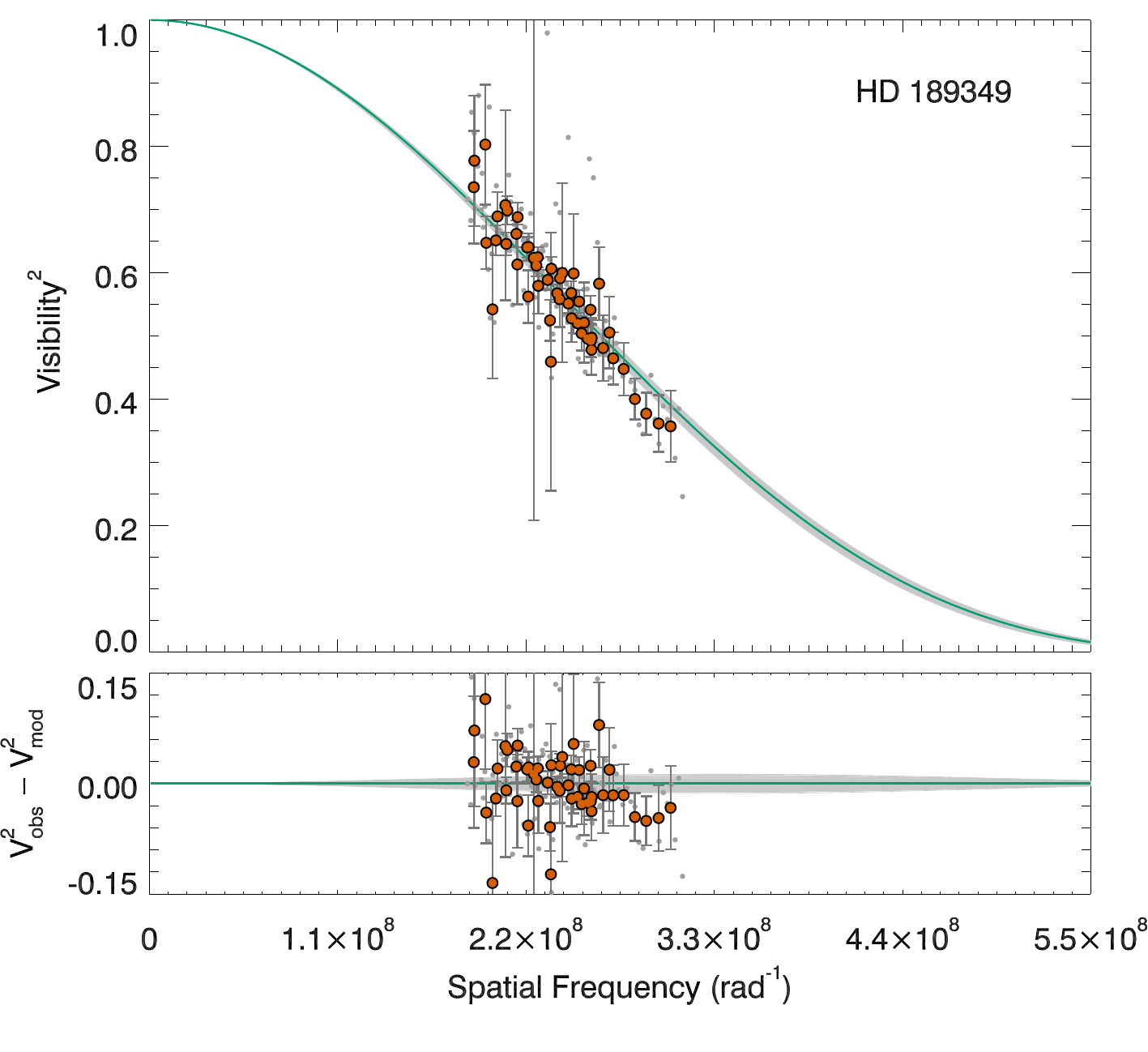}    
      \caption{Squared visibility vs. spatial frequency for HD\,188512 and
HD\,189349. The lower panel shows the residuals from the fit.
The error bars have been scaled to the reduced $\chi^{2}$. For HD\,188512,
the reduced $\chi^{2}$\,=\,2.6 
and for HD\,189349
$\chi^{2}$\,=\,1.5. 
All lines and
symbols are the same as for Fig.\ref{figures1}. }
         \label{figures4}
   \end{figure}

\subsection{Modelling of limb-darkened angular diameters}
\label{models}

We fitted the calibrated fringe visibilities with a limb-darkened disc model. It is common in interferometric studies for a linearly limb-darkened disc model to be used, however,  as in our previous studies \citep{karovicova18,karovicova20}, here we use the four-term non-linear limb-darkening law of \citet{claret00}:
\begin{equation}
    \frac{I\left(\mu\right)}{I\left(1\right)} = 1 - \sum_{k}^{4}a_k\left(1-\mu^{k/2}\right).\label{eqn:LD-4Term}
\end{equation}
Here, $I(\mu)$ is the intensity of the stellar disc at $\mu = \cos(\gamma)$, $\gamma$ is the angle between the line-of-sight and the normal to a given point on the stellar surface, and $a_k$ are the limb-darkening coefficients. 

We use this higher order limb-darkening law because it provides a better fit to the centre-to-limb intensity profiles predicted by model atmospheres \citep[e.g.][]{claret00,magic15}. A linear limb-darkening law, by contrast, tends to produce stronger limb-darkening across most of the disc than is warranted because it is not able to fit both the precipitous drop in intensity towards the limb and the more gradual decrease in intensity elsewhere.

Following the fringe visibility for a generalized polynomial limb-darkening law \citep{quirrenbach96}, the fringe visibilities for the four-term non-linear limb-darkening model are given by
\begin{multline}
    V = \left(\frac{1-\sum a_k}{2}+\sum_k\frac{2a_k}{k+4}\right)^{-1} \times \\
 \left[\left(1-\sum_k a_k\right)\frac{J_1(x)}{x}+\sum_k a_k 2^{k/4}\Gamma\left(\frac{k}{4}+1\right)\frac{J_{k/4+1}(x)}{x^{k/4+1}} \right],
 \end{multline}
where $x=\pi B \theta \lambda^{-1}$, with $B$ the projected baseline, $\theta$ the angular diameter, $\Gamma(z)$ is the gamma function, and $J_n(x)$ is the $n$th-order Bessel function of the first kind. The quantity $B\lambda^{-1}$ is the spatial frequency.

As for our previous studies, we determined the limb-darkening coefficients from the STAGGER grid of ab initio 3D hydrodynamic stellar model atmosphere simulations \citep{magic13}. We determined the coefficients for each model in the grid by fitting Eq.~\ref{eqn:LD-4Term} to the $\mu$-dependent synthetic fluxes calculated by \citet{magic15} in each of the 38 wavelength channels of PAVO. We interpolated within this grid to determine the coefficients for each star. These limb-darkening coefficients are given in Tables A.1. through A.7., available at the CDS. In the appendix, we include one table for one of the stars as an example.
For ease of comparison with previous studies, we also provide angular diameter values determined using a linear limb-darkened law, with coefficients determined from the grid of \citet{claret11} in Table~\ref{table:linear}.

\section{Methods}

\subsection{Bolometric flux}
\label{bol_flux}

Bolometric fluxes and associated uncertainties were derived with the exact same procedure described in \cite{karovicova20}. Briefly, we adopted an iterative procedure to interpolate over the tables of bolometric corrections\footnote{\url{https://github.com/casaluca/bolometric-corrections}} of \cite{Casagrande_VandenBerg14,Casagrande_VandenBerg18a}. We used Hipparcos $H_p$ and Tycho2 $B_TV_T$ magnitudes for all stars, and 2MASS $JHK_S$ only when the quality flag was `A'. The adopted bolometric corrections are listed in Table~\ref{tab:bolcor}, and already take reddening into account for stars affected by it (see Table \ref{tab:parameters} for the adopted E(B-V) values). 


Uncertainties in bolometric fluxes are obtained running a Monte Carlo simulation and taking into account the variance of different bands. Uncertainties for all stars amount to $\sim 2$\% at most, with the exception of HD\,189349, which reaches almost 9\%. This uncertainty is driven by the scatter of bolometric corrections in different filters. Whether this is indicative of model inaccuracies, a wrong reddening estimate, or both is not possible to say, but it raises a warning about the feasibility of using this star as a benchmark. 

While we do not take into account inaccuracies in model fluxes, a comparison with absolute spectrophotometry indicates that by using multiple bands, bolometric fluxes from the CALSPEC library can be recovered at a 1\% level for FG stars \citep{Casagrande_VandenBerg18a}. Our sample, however, comprises cooler stars, for which the performances of our bolometric corrections are much less tested \citep[see e.g., discussions in][]{white18,rains21,tayar20}. An encouraging comparison with the absolute spectrophotometry of a few GK subgiants in \citet{white18} indicates that reliable fluxes can still be obtained from our bolometric corrections. Also, a given percentage change in bolometric flux carries a percentage change in effective temperatures that is four times smaller  .

\subsection{Stellar evolution models}
We used the ELLI package\footnote{Available online at \url{https://github.com/dotbot2000/elli}} \citep{Lin18} together with Dartmouth stellar evolution tracks \citep{Dotter2008} to determine stellar masses. The stellar evolution models were computed with alpha enhancement for the metal-poor stars, and are truncated at the RGB tip. 
The code produces an initial guess from a maximum likelihood estimate when comparing isochrones to the observed $T_\mathrm{eff}$, $\log L/L_\odot$ and $[{\rm Fe }/{\rm H}]$. 
A Markov chain Monte Carlo (MCMC) method is then used to sample the posterior distribution of the stellar mass and age, assuming that errors on the observed quantities are uncorrelated. We take the mean and standard deviation of this distribution as our mass estimate. We then compute the surface gravity from the fundamental relation, rewritten in the form: 
\begin{equation}
\log g = \log \frac {G M}{R^2} = \log \frac {4 G M \varpi^2}{\theta^2},
\label{eq:logg}
\end{equation}
where $G$ is the gravitational constant and $\varpi$ is the parallax.

For three stars, namely, HD~175955, HD~189349, and HD~185351, the MCMC approach produced a posterior distribution that did not match the observed parameters well. For two of the stars, HD~175955 and HD~185351, we instead selected the mass that best reproduced the observations, in the sense that it maximized the likelihood function. We adopted a representative error estimate of 10\,\% for these stars.
The third star, HD~189349, is unusually warm relative to other red giant branch stars of similar luminosity, indicating a very low age (or high mass). Indeed, asteroseismology indicates that this star is undergoing core helium burning (HeB) and thus belongs to the red clump (RC) rather than the red giant branch
\citep{takeda15}. 
Comparison to horizontal branch (HB) evolution tracks yields a fair match for $M = 0.9 \pm 0.1\,\text M_\odot$, as opposed to the RGB solution $M = 2.1 \pm 0.2\,\text M_\odot$. 
%
For this star, we opt instead for the asteroseismic surface gravity: adopting $\nu_\text{max} = 29.03 \pm 0.52\,\mu \text{Hz}$ from \citet{yu18}, we derive $\log g = 2.388 \pm 0.009$ using the scaling relation
\begin{equation}
    \frac{g}{g_\odot} = \frac {\nu_\text{max}}{\nu_{\text{max},\odot}}  \left( \frac{T_\text{eff}}{T_{\text{eff},\odot}} \right)^{3/2}.
\end{equation}
For completeness, we also use a scaling relation 
from \citet{Kjeldsen95},
\begin{equation}
    \frac{M}{M_\odot} = \frac {\nu_\text{max}}{\nu_{\text{max},\odot}}  \sqrt{ \frac{T_\text{eff}}{T_{\text{eff},\odot}}}  \left( \frac{R}{R_\odot} \right)^2
,\end{equation}
to derive the seismic mass, $M = 0.728 \pm 0.025$, but we note that the surface gravity for this star has been derived directly, yielding smaller error bars thanks to the cancellations of terms.


\subsection{Spectroscopic analysis}

High-resolution spectra for the stars were extracted from the 
ELODIE \citep[$R \approx 42\,000$,][]{Moultaka2004} and 
FIES \citep[$R \approx 65\,000$,][]{Telting2014} archives. 
We determined the stellar iron abundances using a custom pipeline based on the spectrum synthesis code SME \citep{piskunov_spectroscopy_2017}
using MARCS model atmospheres \citep{gustafsson_grid_2008}
and pre-computed non-LTE departure coefficients for Fe \citep{amarsi_non-lte_2016-1}.

We selected unblended lines of \ion{Fe}i and \ion{Fe}{ii} between 4400 and 6800\,\AA\ with 
accurately known oscillator strengths from laboratory measurements. 
For saturated lines, we ensured that collisional broadening parameters were available from ABO theory \citep{barklem_list_2000,barklem_broadening_2005}.
Abundances were also measured from solar spectra recorded with the same spectrographs as our target stars, based on observations of light reflected off the Moon (ELODIE) and Vesta (FIES). We thereby produce solar-differential abundances, which mostly cancels uncertainties in oscillator strengths as well as potential systematic differences between the spectrographs.
We estimated the iron abundance of each star from the outlier-resistant mean with $3 \sigma$ clipping.
We also report the difference in abundance between lines of \ion{Fe}i and \ion{Fe}{ii} as an estimate of how closely our fundamental stellar parameters fulfill the ionization equilibrium.
Finally, we compute a systematic uncertainty on the metallicity, which we derive by perturbing the input parameters one at a time according to their formal errors, and we add these differences in quadrature.

\section{Results and discussion}
\label{results}
\subsection{Recommended stellar parameters}

In this section, we present angular diameters and fundamental stellar parameters 
for a set of seven giant and subgiant stars: 
HD\,121370 ($\eta$\,Boo), HD\,161797 ($\mu\,$Her), HD\,175955, HD\,182736, HD\,185351, HD\,188512 ($\beta$\,Aql), and HD\,189349.
Six of these stars have been recommended
as benchmark stars.
The radius and mass are estimated from measurements of $\theta_{LD}$, $F_\mathrm{bol}$ and parallax.
All the values are summarised, along with their luminosity, 
in Table~\ref{der_parameters}.
The final fundamental stellar parameters of  $T_\mathrm{eff}$, $\log\,g$, and $[\mathrm{Fe/H}]$ are listed in Table~\ref{der_parameters2} and also discussed below.

\subsection{Uncertainties}
\label{uncertainties}

The desired precision
requested by the spectroscopic
teams such as {\it Gaia}-ESO or GALAH is around 1\% (or around 40-60~K). We show that for the stars in our sample the uncertainties in $T_\text{eff}$ are less than 50~K (or less than\,1\,\%), with the exception of HD\,189349, which we do not recommend as a benchmark. 
The uncertainty for this star is 2\% and it is
dominated by uncertainties in bolometric flux. 
Overall, the errors in $T_\text{eff}$ for stars in our sample resulting from uncertainties in the limb-darkened angular diameters are at most 31\,K, with a median of just 24\,K (0.5\,\%).

The final uncertainties in $T_\mathrm{eff}$ 
consist of the uncertainties in the bolometric flux and the uncertainties in the angular diameter.
We present the contributions from each measurement independently in Table~\ref{errors}. For clarity, the dominating uncertainty is highlighted in boldface. 
The presented uncertainties have been computed by artificially setting the uncertainties from the other measurement to zero. 
We note that in the final error estimates of  $T_{\mathrm{eff}}$,
we propagate the statistical measurement uncertainties in 
$\log\,g$ and $\mathrm{[Fe/H]}$ from the
isochrone fitting and spectroscopic analysis and fold them into the uncertainties
in the angular diameters. The median uncertainties in 
$\log\,g$ and $\mathrm{[Fe/H]}$ across our sample of
stars are 0.034\,dex and 0.07\,dex, respectively (for details, see Table~\ref{der_parameters2}).

   \begin{table*}
             \caption{{Uncertainties in $T_\mathrm{eff}$ and how they propagate
       from the underlying measurements}}
  \centering   
  \begin{tabular}{l l r r  r r }      
\hline\hline   
Star &   $T_\mathrm{eff}$  & e$T_\mathrm{eff}$ & e$T_\mathrm{eff}$  & e$F_\mathrm{bol}$ $^{a}$ & e$\varTheta_{LD}$ $^{b}$ \\
    &      (K) & (K)& (\%) & (K) & (K)\\ 
  \hline 
HD\,121370 & 6090   & 29  &  0.5 & 15      &  \bf{25} \\
HD\,161797 & 5596   & 22  & 0.4  & 9       &  \bf{21} \\
HD\,175955 & 4568   & 42  & 0.9  & 28      &  \bf{31} \\
HD\,182736 & 5229   & 37  & 0.7  & \bf{29} &  24      \\
HD\,185351 & 5025   & 22  & 0.4  & 8       &  \bf{20} \\
HD\,188512 & 5113   & 20  & 0.4  & 9       &  \bf{17} \\
HD\,189349 & 5199   & 116 & 2.2  & \bf{112}&  31      \\
   \hline  
\end{tabular}
\tablefoot{\tablefoottext{a}{The uncertainties contribution from the bolometric flux if the
$\varTheta_{LD}$ uncertainties are set to 0.
$^{(b)}$  The uncertainties arising entirely from the angular diameter measurements if the $F_\mathrm{bol}$
 uncertainties are set to 0. The dominating uncertainty is highlighted in boldface.}}
\label{errors}
  \end{table*}

\subsection{Comparison with angular diameter values in the literature}
\label{comparison}

\begin{figure}
        \includegraphics[width=\columnwidth]{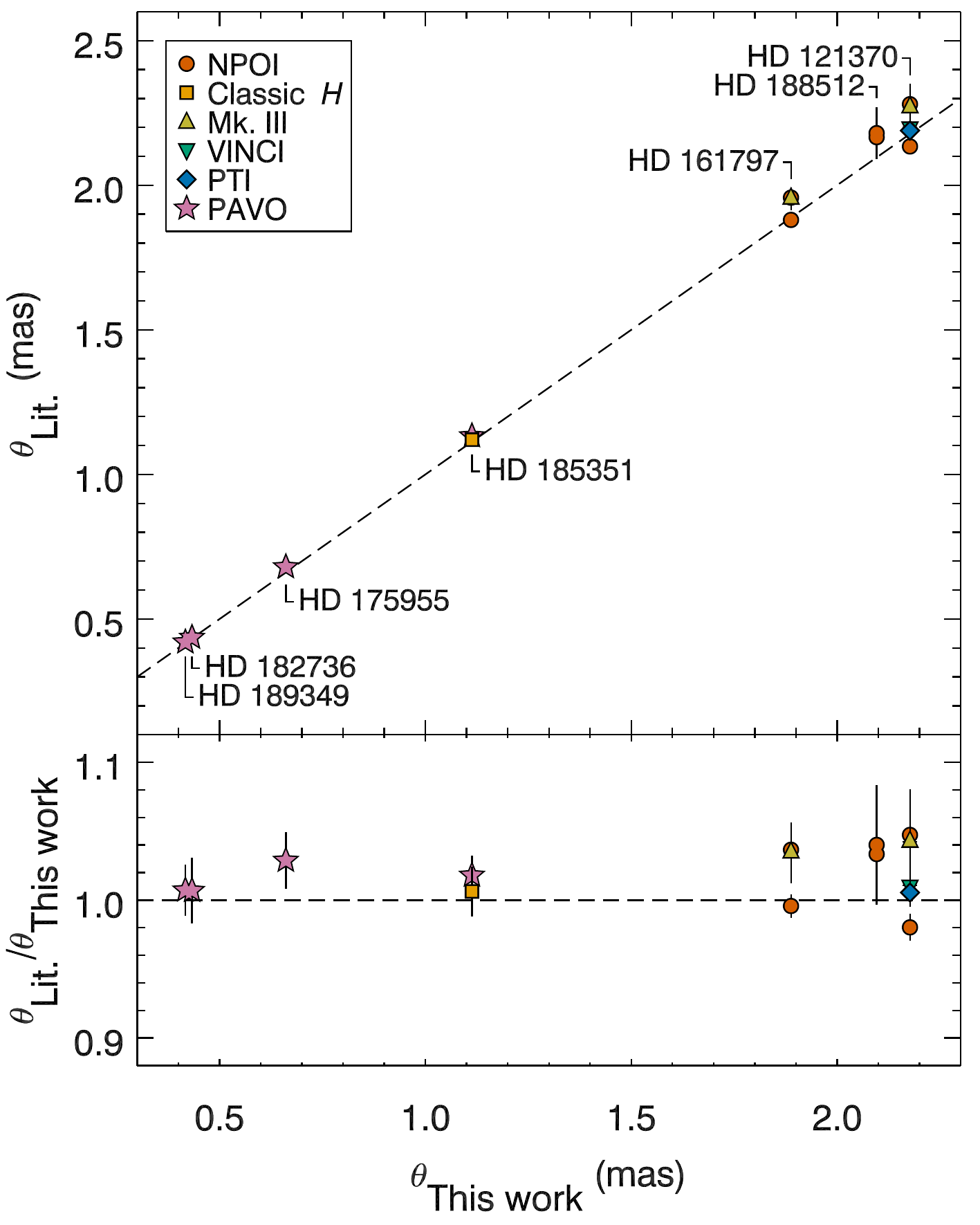}
    \caption{Comparison of limb-darkened angular diameters from the literature with measurements from this work. Symbols correspond to the beam combiner used for the literature measurement: NPOI (red circle), CHARA Classic in the $H$ band (orange square), Mark III (yellow triangle), VLTI/VINCI (green triangle), PTI (blue diamond), and PAVO (pink stars).}
    \label{fig:compare}
\end{figure}

All stars in our sample have been previously interferometrically observed and their angular diameters reported in the literature.
We conducted new observations for some of the stars. In addition, for all stars we carried out a fresh data reduction and analysis of the data applying our updated analysis and updated treatment of limb-darkening.
We list our measurements of angular diameters, $\theta_{LD}$, together with values reported in the literature in Table~\ref{prediam}. We compare the presented values in Fig.~\ref{fig:compare}.\\

{\bf HD\,121370 ($\eta$\,Boo)} We conducted new observations of this star and measured angular diameter as
$\theta_{LD}=2.173\pm0.018\,\text{mas}$.
The star was previously observed  using various interferometric instruments in several studies (see Table~\ref{prediam}). This target is already in use as a benchmark. 
The Gaia-ESO benchmark team \citep{jofre15} adopted the angular diameter measured by \citet{vanBelle07}. Our value is in agreement within the uncertainties with this value. The largest disagreement between our $\theta_{LD}$ and the other measurements is found for the Mk. III measurement by \citet{mozurkewich03}, which differs by 3$\sigma$.

{\bf HD\,161797 ($\mu$\,Her)} We measured the angular diameter of this star as $\theta_{LD}=1.888\pm0.0014\,\text{mas}$.
This star was previously observed by \citet{mozurkewich03} with the Mk.\,III interferometer, who reported a value of
$\theta_{LD}=1.953\pm0.0039\,\text{mas}$.
It has also observed with NPOI by \citet{baines14}, who found $\theta_{LD}=1.957\pm0.0012\,\text{mas}$, and \citet{baines18}, who found $\theta_{LD}=1.880\pm0.0008\,\text{mas}$. Our measurement agrees with the latter measurement. This star was suggested as a possible benchmark star in \citet{heiter15}.

{\bf HD\,175955} For this star, we report
$\theta_{LD}=0.661\pm0.009\,\text{mas}$.
This data was previously observed by \citet{huber12} with the PAVO instrument.
These authors reported $\theta_{LD}=0.680\pm0.01\,\text{mas}$, having fitted the data with a linearly limb-darkened disc model, with the limb-darkening coefficient determined from the grid calculated by \citet{claret11} based on 1D ATLAS models.
This star has also been suggested as a possible benchmark star.
Therefore, for consistency, we carried out a fresh data reduction and reanalysed this data with limb-darkening coefficients from 3D model atmospheres, using a higher-order limb-darkening model.
In this case, our angular diameter differs in comparison to \citet{huber12} by 1.4$\sigma$.  This can be attributed with
the changes in our fresh data reduction. We discuss the impact of the fresh reduction of the PAVO data in more detail below.

{\bf HD\,182736} We measured the angular diameter of this star as $\theta_{LD}=0.433\pm0.0009\,\text{mas}$. The same data was also previously analysed by \citet{huber12},
who reported $\theta_{LD}=0.436\pm0.005\,\text{mas}$, which agrees with our value.
Because this star was suggested as a possible benchmark star, we reanalysed the data applying limb-darkening coefficients from 3D model atmospheres and using a higher-order limb-darkening model.

{\bf HD\,185351} For this star, we report
$\theta_{LD}=1.113\pm0.009\,\text{mas}$.
We again reanalysed PAVO data observed by \citet{johnson14}, who reported
$\theta_{LD}=1.133\pm0.0013\,\text{mas}$.
\citet{johnson14} also reported observations with the Classic\,$H$ instrument at CHARA, finding $\theta_{LD}=0.120\pm0.018\,\text{mas}$.
Our value agrees within uncertainties with the Classic\,$H$ measurement, but differs from the measurement based on the PAVO data by 1.3$\sigma$. Again, this can be attributed with the changes in our fresh data reduction, discussed in more detail below.

{\bf HD\,188512 ($\beta$\,Aql)} We new observations of this star and report an angular diameter of $\theta_{LD}=2.096\pm0.014\,\text{mas}$. 
This star was previously
observed with NPOI by \citet{nordgren99} and \citet{baines14}.
\citet{nordgren99} reported 
$\theta_{LD}=2.18\pm0.09\,\text{mas}$
and \citet{baines14} reported
$\theta_{LD}=2.166\pm0.009\,\text{mas}$.
Our measurement agrees with that of \citet{nordgren99}, although that measurement has a large uncertainty. There is, however, a 4.2$\sigma$ disagreement with the value found by \citet{baines14}.
For this star, we collected very high quality data, fully resolving the star and covering the visibility curve into the second lobe, thus offering a high level of reliability to our measurement.
A detailed analysis of the data covering the second lobe of the visibility function is beyond the scope of this study and will be addressed in a later paper. This star has been suggested as a possible benchmark star \citep{heiter15}.

{\bf HD\,189349} We measured the angular diameter of this star to be $\theta_{LD}=0.417\pm0.005\,\text{mas}$. These PAVO data were previously analysed by \citet{huber12},
who reported $\theta_{LD}=0.420\pm0.006\,\text{mas}$.
Because the star was suggested as a possible benchmark star, we reanalysed it here for consistency. We find that the angular diameter measurements are in good agreement. While this star was recommended as a benchmark by \citet{heiter15}, at this moment we do not recommend this star as a potential benchmark due to its uncertain bolometric flux and, subsequently, its uncertain $T_\mathrm{eff}$ values (see Section~\ref{bol_flux}).\\
\\
\\
Our fresh reduction of the previously published PAVO data had only a minimal impact on the determined angular diameters. Potential differences in the reductions that could have had an impact include making different choices for data cuts based on signal-to-noise ratio (S/N) and other metrics, while also using new calculations of calibrator sizes. In some cases, our choice not to use the $t_0$ correction (as described in Sect.~\ref{sec:observations}) can have a more significant impact. Additionally, we used all 38 of the PAVO wavelength channels, whereas the previous studies only used the central 23 channels between 650 and 800\,nm. The effect of the these data reduction changes alone can be determined by comparing the linear limb-darkened diameters given in Table~\ref{table:linear} with the diameters reported in the literature. The diameters in Table~\ref{table:linear} agree (within 1$\sigma)$ with the literature values.

The remaining difference between our final adopted angular diameters and the literature values for these targets is attributed to the updated limb-darkening treatment. The previous PAVO studies fitted the data using a linear limb-darkening model, with coefficients calculated from 1D ATLAS model atmospheres by \citet{claret11}. A single coefficient for the $R$-band was used across all wavelength channels. As discussed in detail by \citet{karovicova21}, the combined effect of this was that the adopted coefficients implied stronger limb-darkening than justified by the model atmospheres. As a consequence, the previous angular diameters were too large and, on average, our final adopted angular diameters are 1.5\% smaller.


   \begin{table*}
   \caption{Prior angular diameters}    
  \centering   
  \begin{tabular}{l c c l l } 
  \hline
  \hline
Star              &             Our value                &      Literature value    &      Reference                &      Instrument\\
&{(mas)}&{(mas)}&&\\
\hline
HD\,121370                &     2.173$\pm$0.018 &&&\\
&& 2.28$\pm$0.07 & \citet{nordgren01} & NPOI \\
&& 2.269$\pm$0.025 & \citet{mozurkewich03} & Mk. III \\
&& 2.200$\pm$0.027 & \citet{thevenin05} & VLTI/VINCI \\
&& 2.189$^{+0.006}_{-0.014}$ & \citet{vanBelle07} & PTI\\
&& 2.134$\pm$0.012 & \citet{baines14} & NPOI \\
\hline
HD\,161797        &             1.888$\pm$0.014 &&&\\
&& 1.953$\pm$0.039 & \citet{mozurkewich03} & Mk. III \\
&& 1.957$\pm$0.012 & \citet{baines14} & NPOI \\
&& 1.880$\pm$0.008 & \citet{baines18} & NPOI\\
\hline
HD\,175955                &     0.661$\pm$0.009  &  &  &                \\              &&         0.68$\pm$0.01                    &      \citet{huber12}  &      CHARA/PAVO\\
\hline
HD\,182736                &     0.433$\pm$0.009 &&&\\
&&      0.436$\pm$0.005                  &      \citet{huber12}  &      CHARA/PAVO\\
\hline
HD\,185351                &     1.113$\pm$0.009 &&&\\
& & 1.120$\pm$0.018              &              \citet{johnson14}        &       CHARA/Classic $H$\\
& & 1.133$\pm$0.013 & \citet{johnson14} & CHARA/PAVO \\
\hline
HD\,188512                &     2.096$\pm$0.014 &&&\\
&& 2.18$\pm$0.09 & \citet{nordgren99} & NPOI\\
&& 2.166$\pm$0.009 & \citet{baines14} & NPOI\\
\hline
HD\,189349                &     0.417$\pm$0.005 &&&\\
&&      0.420$\pm$0.006                  &      \citet{huber12}  &      CHARA/PAVO\\

   \hline  
   \label{prediam}
\end{tabular}
  \end{table*}

\subsection{Spectroscopic analysis}
\begin{figure}
\includegraphics[width=\hsize]{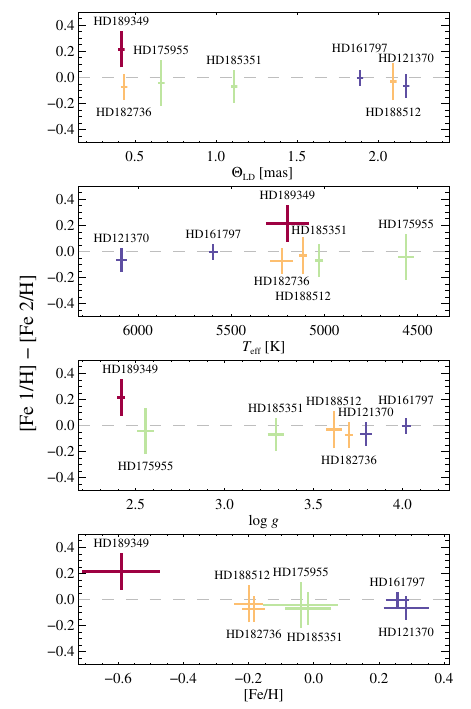}
\caption{Deviations from ionisation equilibrium, that is,\ the difference between the abundances determined from lines of neutral and ionised iron as a function of the measured stellar parameters. Vertical and horizontal lines represent the combined uncertainties from the two measurements. Stars are colour-coded according to metallicity, as shown in the bottom plot. }
\label{fig:ionisation_equilibrium}
\end{figure}

Our iron abundance determinations from lines of neutral and ionised iron yield very good agreement, as illustrated in Fig.~\ref{fig:ionisation_equilibrium}. 
The abundance differences are small, consistent with zero, for all stars except HD 189349. This star, which has the lowest metallicity in the sample, and is also the only core helium-burning star, shows a positive deviation of 0.2\,dex. 

Errors in stellar parameters have different impact on lines of neutral and singly ionized iron. The sensitivity is such that a deviation from ionization equilibrium of 0.05\,dex may correspond to an error of approximately 50\,K in $T_\text{eff}$, or 0.10\,dex in $\log\,g$. The excellent agreement for all but one stars in our sample therefore indicate that our error estimates are realistic, and that no significant systematics are present. For HD 189349, the larger deviation by 0.2\,dex may indicate either an error in its temperature by upwards of 200\,K, an error in $\log\,g$ by as much as 0.4\,dex, or issues related to spectroscopic modelling \citep[see e.g.][]{Lebzelter2012}.

\subsection{Comparison with asteroseismology}

Each of our targets exhibits solar-like oscillations that are stochastically excited and damped by convection. Measurements of the oscillation frequencies can place tight constraints on the interior structure of a star, and can be used to infer stellar properties including the radius. Here, we briefly compare our interferometric radii with radii derived from asteroseismic studies in the literature.

Three of our targets have been the subject of ground-based radial velocity campaigns to detect their oscillations: HD\,121370 \citep[$\eta$\,Boo;][]{kjeldsen03,carrier05}, HD\,161797 \citep[$\mu$\,Her;][]{bonanno08,grundahl17}, and HD\,188512 \citep[$\beta$\,Aql;][]{corsaro12}. The other four stars were observed by the Kepler mission. HD\,175955, HD\,182736, and HD\,189349 were studied by \citet{huber12}. Additionally, HD\,175955 and HD\,189349 were amongst the 16,000 red giants in the sample of \citet{yu18}. Also, HD\,185351 was one of the brightest stars observed during the nominal Kepler mission and its oscillations were studied by \citet{johnson14}.

The simplest way to determine masses and radii from asteroseismic measurements is to exploit scaling relations for two quantities that parameterise the observed oscillation spectrum. The frequency at which the oscillation amplitude peaks, $\nu_\mathrm{max}$, is observed to scale as follows \citep{Brown91,Kjeldsen95}:
\begin{equation}
    \frac{\nu_\mathrm{max}}{\nu_\mathrm{max,\odot}} \approx \left(\frac{M}{\mathrm{M_\odot}}\right)\left(\frac{R}{\mathrm{R_\odot}}\right)^{-2}\left(\frac{T_\mathrm{eff}}{\mathrm{T_{eff,\odot}}}\right)^{-1/2}.
\end{equation}
The large frequency separation, $\Delta\nu$, between modes of the same angular degree and consecutive radial order scales as \citep{ulrich86}:
\begin{equation}
    \frac{\Delta\nu}{\Delta\nu_{\odot}} \approx \left(\frac{M}{\mathrm{M_\odot}}\right)^{1/2}\left(\frac{R}{\mathrm{R_\odot}}\right)^{-3/2}.
\end{equation}

Combining these two relations with the known $T_\mathrm{eff}$ allows the mass and radius to be determined directly \citep[e.g.][]{Kallinger10}. Comparisons with independent measurements have found radii determined from the scaling relation to be accurate to $\sim5\%$ \citep[e.g][]{huber12,white13,huber17,brogaard18}. However, as the scaling relations rely on a number of assumptions, including that the stars are homologous to the Sun, deviations arise for stars that have significantly different structures, such as red giants. These deviations have been investigated, particularly for the $\Delta\nu$ scaling relation, which has a more established theoretical basis, and corrections to improve the relations have been proposed \citep[e.g.][]{White11,guggenberger16,sharma16,rodrigues17,viani17}.

\begin{figure}
        \includegraphics[width=\columnwidth]{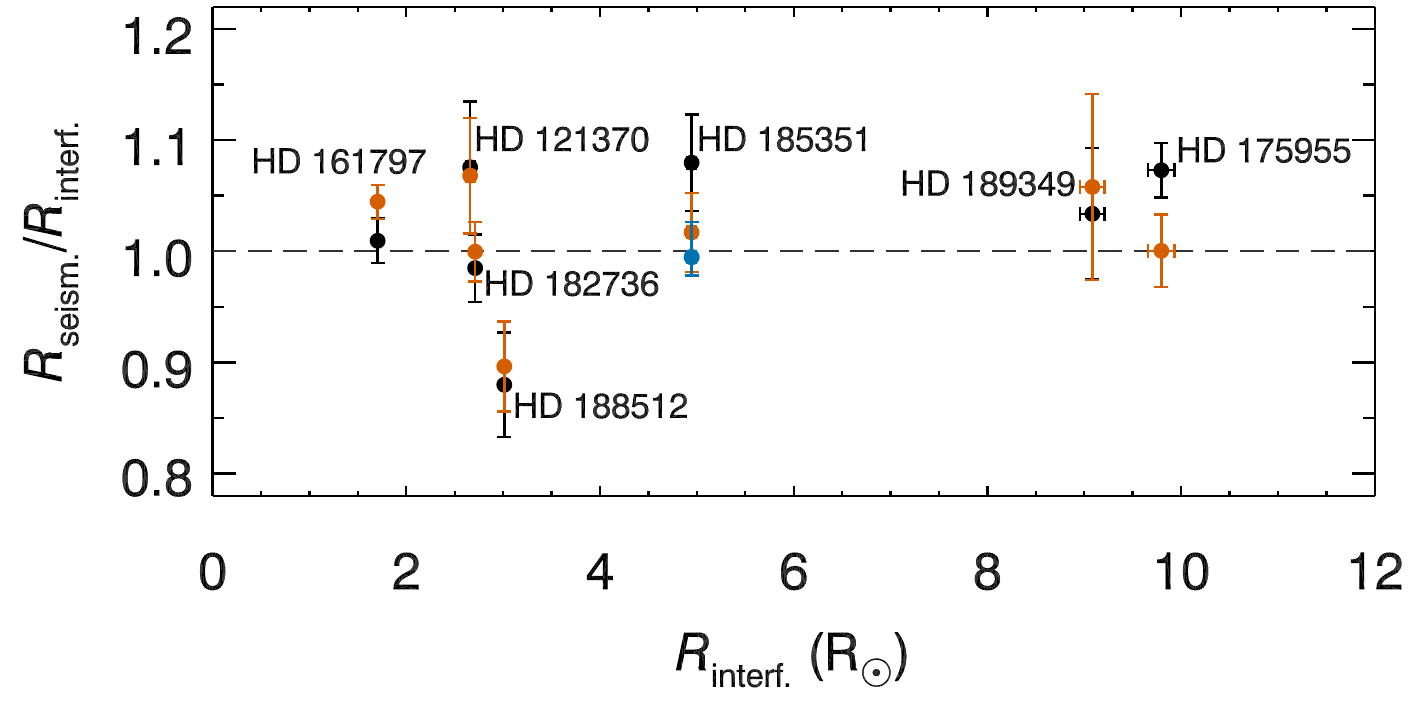}
    \caption{Comparison of radii determined from asteroseismic scaling relations (uncorrected in black, corrected in orange) and modelling (blue) with interferometric radii from this work.}
    \label{fig:seismic}
\end{figure}

Figure~\ref{fig:seismic} shows the scaling relation radii compared to the interferometric radii for our sample. The uncorrected scaling relation radii are typically accurate to better than 10\% for this sample. The scaling relation radii for HD\,121370 and HD\,188512 are relatively uncertain, owing to the comparatively short length and lower signal-to-noise ratio of the ground-based data for these stars. Two of the giants observed by Kepler, HD\,185351 and HD\,175955, have an uncorrected scaling relation radius that disagrees with the interferometric radius, the latter by 3$\sigma$. The scaling relation radii corrected using the $\Delta\nu$ corrections of \citet{sharma16} show better agreement with the interferometric radii in general, and for these two stars, in particular. This provides further support for the use of such corrections. It is only for one star, HD\,161797, that the correction does indeed worsen the agreement with the interferometric radius significantly.

A more rigorous method of determining stellar parameters from asteroseismology than using scaling relations is detailed stellar modelling using the observed oscillation frequencies. Three of these stars have been the subject of such efforts. As one of the first stars to have solar-like oscillations detected, HD\,121370, has been studied in detail \citep[e.g.][]{christensen-dalsgaard95,DiMauro03,carrier05,Straka06}. However, these studies did not report a final value for the radius with uncertainties. The radius of the best fitting model found by \citet{carrier05} is 2.72\,$\mathrm{R_\odot}$, which is in better agreement with the interferometric radius than the scaling relation radii. \citet{li19} modelled HD\,161797, however, they used the interferometric radius as one of the constraints on their models, so a valid comparison cannot be made with their results. Finally, HD\,185351 was modelled by  \citet{hjoerringgaard17}, who sought to resolve tension between the interferometric and asteroseismic constraints by varying the input physics of the models; in particular, the convective core overshooting and mixing length parameter. Their radius, shown in Fig.~\ref{fig:seismic}, agrees well with our interferometric radius. More detailed modelling of more stars with both asteroseismic and interferometric constraints ought to allow for the calibration of the input physics of models, such as convection \citep[e.g.][]{Joyce18}.

\section{Conclusions}

We presented a sample of giant and subgiant stars with highly accurate and reliable fundamental 
stellar parameters, extending the sample of stars that may be used as benchmarks
for large stellar surveys.
This is the third in a series of papers with this scientific goal. The sample in this paper consists of 
HD\,121370 ($\eta$\,Boo), HD\,161797 ($\mu\,$Her), HD\,175955, HD\,182736, HD\,185351, HD\,188512 ($\beta$\,Aql), and HD\,189349. One of these stars, HD\,121370, is already listed as a Gaia-ESO benchmark star and we revised its stellar parameters here. A further five stars have also been proposed as possible benchmark stars by \citet{heiter15}.


We determined angular diameters for the targets using data from the PAVO beam combiner at the CHARA Array. For three of the stars, we present new observations. For the other four stars, we reanalysed previous observations to ensure the stellar parameters were determined consistently. Limb-darkening coefficients were determined from the STAGGER grid of 3D hydrodynamic stellar model atmospheres.

We computed bolometric fluxes from multi-band photometry, interpolating iteratively on a grid of synthetic stellar fluxes to ensure consistency with the final adopted
stellar parameters. Effective temperatures were determined directly from the angular diameters and bolometric fluxes.

We determined $\mathrm{[Fe/H]}$ based on high-resolution spectroscopy. We used isochrone fitting to derive mass, and parallax measurements to constrain the absolute luminosity. After iterative refinement between interferometry, spectroscopy and photometry we derived the final fundamental parameters of $T_\mathrm{eff}$, $\log\,g$, and $\mathrm{[Fe/H]}$ for all stars. 

For six of our seven targets, we reached the desired precision of $\lessapprox$\,1\% in $T_\mathrm{eff}$. Only one star (HD\,189349) showed a somewhat larger $T_\mathrm{eff}$ uncertainty of 2\,\%.
The $T_\mathrm{eff}$ uncertainty for this star is dominated by uncertainty in the bolometric flux. We do not recommend that this star be used as a benchmark until its bolometric flux can be more reliably measured. For the surface gravity, $\log\,g,$ we reached a median precision of 0.034\,dex 
and for metallicity $\mathrm{[Fe/H],}$ we reached a median precision of 0.07\,dex.

The giant and subgiant stars presented here, in conjunction with our previous results for metal-poor \citep{karovicova20} and dwarf stars \citep{karovicova21}, form a sample of benchmark stars with consistently derived, highly reliable fundamental stellar parameters. The precision we have been able to achieve is essential to a consistent and reliable calibration of atmospheric parameters across large spectroscopic surveys. Consequently, the correctly determined atmospheric parameters will help us
to better understand the evolution of stars in the Milky Way.\\

\begin{acknowledgements}
      I.K. acknowledges the German
      \emph{Deut\-sche For\-schungs\-ge\-mein\-schaft, DFG\/} project
      number KA4055 and the European Science Foundation - GREAT Gaia Research for European Astronomy Training.
      This work is based upon observations obtained with the Georgia State University Center for High Angular Resolution Astronomy Array at Mount Wilson Observatory. The CHARA Array is supported by the National Science Foundation under Grants No. AST-1211929 and AST-1411654. Institutional support has been provided from the GSU College of Arts and Sciences and the GSU Office of the Vice President for Research and Economic Development.
      Funding for the Stellar Astrophysics Centre is provided by The Danish National Research Foundation.
      L.C. is the recipient of the ARC Future Fellowship FT160100402. T.N. acknowledges funding from the Australian Research Council (grant DP150100250).
      Parts of this research were conducted by the Australian Research Council Centre of Excellence for All Sky Astrophysics in 3 Dimensions (ASTRO 3D), through project number CE170100013.
      D.H. acknowledges support from the Alfred P. Sloan Foundation, the National Aeronautics and Space Administration (80NSSC19K0379), and the National Science Foundation (AST-1717000).
      This work is based on spectral data retrieved from the ELODIE archive at Observatoire de Haute-Provence (OHP). 
\end{acknowledgements}

%
%

\bibliographystyle{aa} 
\bibliography{ref}

\begin{appendix}
\section{Limb-darkening coefficients in 38 channels.}
\begin{table}[h!]\small
\caption{Limb-darkening coefficients in 38 channels. We show one table for the star HD\,121370 as an example.
Limb-darkening coefficients for the rest of the stars are available at CDS in tables A.1.--A.7.}   \label{table:4}      
\centering                          
\begin{tabular}{llllll}      
\hline\hline  
\multicolumn{5}{l}{HD\,121370}  \\
&\multicolumn{4}{c}{four-term limb-darkening$^a$} \\   
chan.& wavelength&        $a_1$ & $a_2$ & $a_3$ & $a_4$          \\
\hline    
1.   &    0.630     &       1.967    &     -3.239    &      3.420    &     -1.258      \\
2.   &    0.635     &       1.967    &     -3.242    &      3.418    &     -1.256      \\
3.   &    0.639     &       1.975    &     -3.255    &      3.422    &     -1.256      \\
4.   &    0.644     &       1.995    &     -3.317    &      3.496    &     -1.290      \\
5.   &    0.649     &       2.004    &     -3.340    &      3.507    &     -1.290      \\
6.   &    0.654     &       2.066    &     -3.434    &      3.546    &     -1.311      \\
7.   &    0.659     &       2.068    &     -3.448    &      3.556    &     -1.311      \\
8.   &    0.664     &       1.989    &     -3.302    &      3.458    &     -1.276      \\
9.   &    0.670     &       1.996    &     -3.329    &      3.485    &     -1.283      \\
10.  &    0.675     &       1.993    &     -3.323    &      3.470    &     -1.276      \\
11.  &    0.680     &       1.997    &     -3.335    &      3.477    &     -1.277      \\
12.  &    0.686     &       2.004    &     -3.359    &      3.501    &     -1.286      \\
13.  &    0.692     &       2.006    &     -3.362    &      3.495    &     -1.282      \\
14.  &    0.698     &       2.015    &     -3.394    &      3.529    &     -1.296      \\
15.  &    0.704     &       2.021    &     -3.417    &      3.552    &     -1.305      \\
16.  &    0.710     &       2.024    &     -3.426    &      3.556    &     -1.307      \\
17.  &    0.716     &       2.040    &     -3.473    &      3.604    &     -1.325      \\
18.  &    0.722     &       2.039    &     -3.476    &      3.604    &     -1.325      \\
19.  &    0.729     &       2.038    &     -3.475    &      3.592    &     -1.316      \\
20.  &    0.736     &       2.041    &     -3.488    &      3.606    &     -1.323      \\
21.  &    0.743     &       2.055    &     -3.534    &      3.649    &     -1.337      \\
22.  &    0.750     &       2.049    &     -3.517    &      3.627    &     -1.330      \\
23.  &    0.756     &       2.045    &     -3.510    &      3.618    &     -1.327      \\
24.  &    0.763     &       2.054    &     -3.539    &      3.643    &     -1.336      \\
25.  &    0.771     &       2.051    &     -3.537    &      3.639    &     -1.334      \\
26.  &    0.778     &       2.058    &     -3.563    &      3.670    &     -1.349      \\
27.  &    0.786     &       2.047    &     -3.533    &      3.626    &     -1.329      \\
28.  &    0.794     &       2.058    &     -3.569    &      3.663    &     -1.343      \\
29.  &    0.802     &       2.055    &     -3.569    &      3.661    &     -1.342      \\
30.  &    0.810     &       2.053    &     -3.568    &      3.658    &     -1.342      \\
31.  &    0.818     &       2.062    &     -3.600    &      3.687    &     -1.352      \\
32.  &    0.827     &       2.053    &     -3.578    &      3.661    &     -1.341      \\
33.  &    0.835     &       2.065    &     -3.617    &      3.700    &     -1.356      \\
34.  &    0.844     &       2.055    &     -3.592    &      3.669    &     -1.344      \\
35.  &    0.853     &       2.065    &     -3.620    &      3.687    &     -1.347      \\
36.  &    0.863     &       2.058    &     -3.609    &      3.673    &     -1.341      \\
37.  &    0.872     &       2.067    &     -3.646    &      3.715    &     -1.361      \\
38.  &    0.881     &       2.060    &     -3.633    &      3.708    &     -1.363      \\
\hline                                 
\end{tabular}
\flushleft $^{a}$ Limb-darkening coefficients derived from the grid of \citet{magic15}; see text for details.
\end{table}

\end{appendix}
\end{document}